\documentclass[a4paper,10pt]{article}
\usepackage{geometry}
\usepackage{color}
\usepackage{url}
\usepackage{graphicx}
\usepackage{lipsum}
\usepackage[export]{adjustbox}
\usepackage[font=footnotesize,labelfont=bf]{caption}
\usepackage{mathtools}
\usepackage[T1]{fontenc}
\usepackage{amsmath}
\usepackage{amssymb}
\usepackage{hyperref}
\usepackage{fancyhdr}
\usepackage[utf8]{inputenc}
\usepackage{authblk}
\usepackage{rotating}
\usepackage{comment}
\usepackage{authblk}
\usepackage{cite}
\usepackage{url}
\usepackage{bbm}
\usepackage{soul}
\usepackage[normalem]{ulem}
\usepackage{tabularx,booktabs}
\newcolumntype{Y}{>{\centering\arraybackslash}X}
\def\code#1{\texttt{#1}}

\title{\textbf{Non-equilibrium time-dependent solution to discrete choice with social interactions}}
\author[1]{James Holehouse\thanks{Correspondence: james.holehouse@ed.ac.uk; (+44) 0793 890 5690; Rm 3.04, CH Waddington Building, Max Born Crescent, Edinburgh, EH9 3BF.}}
\author[2]{Hector Pollitt}
\affil[1]{School of Biological Sciences, University of Edinburgh,}
\affil[2]{The World Bank.}
\date{\today}

\begin{document}

\maketitle

\abstract

We solve the binary decision model of Brock and Durlauf \cite{brock2001discrete} \textit{in time} using a method reliant on the resolvent of the master operator of the stochastic process. Our solution is valid when not at equilibrium and can be used to exemplify path-dependent behaviours of the binary decision model. The solution is computationally fast and is indistinguishable from Monte Carlo simulation. Well-known metastable effects are observed in regions of the model’s parameter space where agent rationality is above a critical value, and we calculate the time scale at which equilibrium is reached from first passage time theory to a much greater approximation than has been previously conducted. In addition to considering selfish agents, who only care to maximise their own utility, we consider altruistic agents who make decisions on the basis of maximising global utility. Curiously, we find that although altruistic agents coalesce more strongly on a particular decision, thereby increasing their utility in the short-term, they are also more prone to being subject to non-optimal metastable regimes as compared to selfish agents. The method used for this solution can be easily extended to other binary decision models, including Kirman's ant model \cite{kirman1993ants}, and under reinterpretation also provides a time-dependent solution to the mean-field Ising model. Finally, we use our time-dependent solution to construct a likelihood function that can be used on non-equilibrium data for model calibration. This is a rare finding, since often calibration in economic agent based models must be done without an explicit likelihood function. From simulated data, we show that even with a well-defined likelihood function, model calibration is difficult unless one has access to data representative of the underlying model.


\section{Introduction}
It has been 20 years since the publication of \textit{discrete choice with social interactions} \cite{brock2001discrete}, which at the time of writing has over two thousand citations. The success of the publication has spawned a multitude of related publications, all with an interest in modelling how \textit{variably rational} economic agents make decisions under exogenous influence in a system with collective endogenous interactions. More explicitly, the model of Brock and Durlauf considers a system of agents where each agent makes a decision \textit{left} or \textit{right}, where their decision is influenced by global influences (affecting all agents) and collective effects relating to conformity forces between the agents (the more agents deciding one way, the stronger the influence on agents deciding the other way to change their minds). This simple but enlightening model has been extended in various directions in recent years, to multiple choice scenarios \cite{borghesi2007songs}, towards models integrating heterogeneous agents in complex networks in the \textit{random field Ising model} (RFIM) \cite{bouchaud2013crises}, and studying hysteresis in economic networks \cite{hosseiny2019hysteresis}. The binary decision model has similarities to other situations beyond binary choice, a famous example being the mean-field analysis done by Kirman et al.~in studies of the Marseille fish market in order to explain the dynamics of how partially-rational, partially-loyal agents choose sellers of fish \cite{weisbuch2000market,kirman2010complex,moran2021ants}. The study of such models is partially motivated by their simplicity compared to more general models, but also since they seem to be able to replicate some real socio-economic phenomena even given their simplicity. 


One of the extensions of the binary decision model is the generalised model of Bouchaud in \cite{bouchaud2013crises}. Therein, he proposed the RFIM as a unifying framework for the study of socio-economic phenomena, which subscribes agents as being heterogeneous (being predisposed to one decision over another), in a complex network of (possibly non-symmetric) interactions, subject to a global zeitgeist which in principle can change in time. This model is incredibly rich and its dynamics in some cases are described by evolving metastable states and long waiting times to reach equilibrium, very similar to known phenomena from the study of \textit{spin glasses} in physics \cite{mezard1987spin}. However, the downsides of the RFIM lie not in its ability to represent near infinite different realisations, but instead that this complexity restricts the possibility of solving such models analytically. This additionally makes model calibration from such models extremely difficult since the models are often made up of tens of parameters describing the probability distributions for the connections between agents, agent heterogeneity and the time-dependence of the zeitgeist. Therefore, although the RFIM can likely provide realistic descriptions of real-world socio-economic phenomena, the binary decision model of Brock and Durlauf is likely to provide the ideal starting point for model calibration and analytics of real data. We note that Brock and Durlauf's model is very closely related to Kirman's model of ant rationality (which is also known as the Moran model \cite{moran1958random}), with the difference being in the choice of the transition rates between the two possible decisions \cite{kirman1993ants,moran2021ants,moran2020schrodinger,farmer2020self}.


One can in fact draw an analogy to the field of Systems Biology, which uses mathematical models to describe the expression of mRNA and proteins from genes in the DNA \cite{roberts2002molecular,schnoerr2017approximation}. In reality, genes interact with proteins produced by other genes (and even their own proteins), which can be visualised as a gene regulatory network made up of many different network patterns known as motifs \cite{milo2002network}. However, for practical reasons, researchers often ignore the complex interactions present in gene regulatory networks and instead describe genes of interest using a more simplified description. This simplified mechanism is known as the \textit{telegraph model}, and it considers each gene as an on/off switch (either it can produce mRNA or it cannot), a model made up of four rate parameters rather than the potentially hundreds that would be necessary to describe the regulatory network of a particular gene \cite{peccoud1995markovian,iyer2009stochasticity,braichenko2021distinguishing}. Experimental gene expression data is then calibrated to these models to give the set of four parameters (for each gene) describing the simplified model \cite{schwanhausser2011global,halpern2015bursty,larsson2019genomic,suter2011mammalian,herbach2017inferring}. Importantly, the distributions of cellular mRNA numbers are well-fit by telegraph models. In the case of binary decisions, although network effects are important for some situations, as Brock and Durlauf \cite{brock2001interactions} and Kirman state \cite{kirman2010complex}: \textit{one should be careful not to overemphasise network effects unless one has an empirical reason to infer their importance}. Hence, in the same way that the telegraph model provides a simplified description of gene expression, the model of Brock and Durlauf \cite{brock2001discrete} provides a simple go-to model with which to classify different binary decision situations via a small number of parameters.

However, for all the applause of the model of Brock and Durlauf, the solution that they provide for the binary choice model \textit{applies only at equilibrium}, whereas it is now widely known that the time to reach equilibrium, for many socio-economic systems of interest, does not occur on an economically relevant time scale \cite{bouchaud2013crises,sornette2014physics}. Hence, to understand binary decision phenomena more deeply one must consider the whole time-evolution of the dynamics from an initial condition to the equilibrium. In particular, model calibration is likely to be greatly perturbed using a likelihood function based on the equilibrium solution for data trajectories that are not a sample of the equilibrium distribution. In this paper we solve for the probability distribution of the mean-field binary decision model of Brock and Durlauf \textit{in time}, and use our solution to investigate metastability, and the requirements on economic data necessary for model calibration. The method used for our solution also solves Kirman's ant model for the case of discrete numbers of agents, complementary to the study of \cite{moran2020schrodinger}, where a time-dependent solution is provided for a continuous number of agents (i.e., the large agent population limit). This analytic solution allows for faster model calibration compared to simulation based approaches, as well as the analytical construction of a likelihood function.

The paper is structured as follows. In Section \ref{sec:econ_agents} we introduce the mean-field binary decision model of Brock and Durlauf starting from the RFIM of Bouchaud \cite{bouchaud2013crises}. In this way the reader can see how the mean-field description comes from applying various assumptions to the RFIM. Section \ref{sec:selfish} describes the situation of a system of selfish agents who make decisions such that they maximise their utility (if they are rational). In Section \ref{sec:altruistic} we extend this analysis to a system of altruistic agents, where now agents are concerned with optimising the global utility. Then in Section \ref{sec:sol} we solve the mean-field binary decision model in time. Throughout Section \ref{sec:metastab} we explore the solution that we have found the metastable phenomena it possesses (known as a lock-in in the economics community). In the process, using a first passage time analysis, we calculate the time scales of these metastable states. In Section \ref{sec:cali} we use our time-dependent solution to construct a likelihood function that can be used for model calibration, and assess the conditions necessary on simulated data to provide accurate calibration. We then compare the results from our model to the standard tenets of neoclassical economics in Section \ref{sec:compNCE}, before concluding our paper in Section \ref{sec:conc}.

\section{Binary decision model}\label{sec:econ_agents}
\subsection{Selfish agents}\label{sec:selfish}
Consider a system of $N$ economic agents where each agent $i$ can choose between two distinct decisions, \textit{left} denoted by $S_i(t) = -1$, and \textit{right} denoted by $S_i(t) = 1$. Left or right could correspond to an infinite number of binary choice questions, for example, to vote either Democrat or Republican, to buy one stock or another, or whether one is participating in the latest fashion trend (in this case yes or no). Depending on the question one is considering, the model parameters making up a binary decision model are likely to be very different. For example, a binary decision model of an American election is unlikely to have the same model parameters as one describing consumers choosing between two brands of similar cereal: in the case of a cereal, the exogenous influence of advertising is more important than endogenous effects, whereas the collective effects are likely more important in an election model. The attribute $S_i(t)$ denotes the decision currently made by agent $i$ at a time $t$, with the state of the entire system at time $t$ being given by the set of choices made by all agents at that time, denoted $\mathcal{S} = \{S_1,S_2,...,S_N\}$. The RFIM introduced by Bouchaud in \cite{bouchaud2013crises} takes into account three main factors, all of which contribute to the influence on agent $i$, $I_i(t)$, which are (using naming conventions inspired by \cite{bouchaud2013crises}):
\begin{enumerate}
    \item[1.] The \textit{personal inclination} (or predisposition) of the agent to favour one decision over the other. The contribution to $I_i(t)$ from personal inclination is $f_i \in [-\infty,\infty]$, where $f_i<0$ means a predisposition to left and $f_i>0$ means a predisposition to right. In principle, these heterogeneities could be time-dependent, but since they are difficult to empirically quantify one generally assumes they are constant. 
    \item[2.] The \textit{zeitgeist}, a global influence equally affecting all agents, such as news reporting or stock trends. Its contribution to $I_i(t)$ is given by $F(t)\in[-\infty,\infty]$, and we assume its time-dependence directly since it is often global effects (e.g., stock market decline or warming global temperature) that change in time.
    \item[3.] \textit{Agent-to-agent interactions}, attractive/repulsive influences for the agents to attract/repel others from their same view. The strengths of these interactions are stored in the matrix $\mathbf{J}$, where $J_{ij}\equiv [\mathbf{J}]_{ij}$ is the contribution of agent $j$ to $I_i(t)$ such that the total contribution of all agents in neighbourhood $\nu_i$ to the influence of agent $i$ is $\sum_{j\in\nu_i}J_{ij}S_j(t)$.
\end{enumerate}
In summary, each agent is subject to an \textit{influence}, $I_i(t)$, which is the sum of these contributions,
\begin{align}\label{eq:info1}
    I_i(\mathcal{S},t) = f_i + F(t) + \sum_{j\in\nu_i}J_{ij}S_j(t).
\end{align}
The importance of the influence is that, where agents are rational, they will aspire to make a decision (left or right) that agrees with the influence they are subjected to. We explore the mechanisms through which agents update their decisions below, based on the introduction of an agent utility function.

In the mean-field case several simplifications can be made. The main assumption is that all agents are equally connected to each other with strength $J$ meaning that each agent only responds to the average global opinion of all agents $m = \sum_i S_i/N$. As has been widely documented, \textit{the mean-field approximation is qualitatively a good approximation for nearest neighbour interactions where $N\gg 1$ and the number of dimensions $d\geq 4$ or in situations where agents are connected to many other agents} \cite{amit1974ginzburg}. We additionally rescale $J$ by the number of agents $J\to J/N$, such that the agent-to-agent interaction contribution to $I_i(\mathcal{S},t)$ is intensive. Finally, we assume that each agent has the same personal inclination $f$, a factor that is absorbed into the definition of the $F(t)$ in Eq.~\eqref{eq:info1}.  Note that even in the absence of personal preference the agents are still heterogeneous in the sense that they make their own decisions \textit{at different times}. Within this set of approximations the information is a shared function amongst all agents and is given by,
\begin{align}
    I_i(\mathcal{S},t)\equiv I(n(t),t) = F(t) + J m(n(t)),
\end{align}
where we reinterpret the definition of $m(n(t)) = (2n(t)-N)/N$, where $n(t)$ is the number of agents whose decision at $t$ is right (with $N-n(t)$ deciding left). For brevity, we will often drop the time-dependence of $n$ on $t$, although it is always implicit. The function $m(n)$, as well as being the average global opinion, is the \textit{order parameter} of the system and takes discrete values in $[-1,1]$ for finite $N$: $m(N) = 1$ if all agents take the right decision, $m(N/2) = 0$ if equal numbers decide right and left, and $m(0)=-1$ if all agents decide left. Hence, $m(n)$ appropriately characterises the state of the system for any value of $N$. 

\begin{figure}[h!]
\captionsetup{width=1.0\textwidth}
\centering
\includegraphics[width=1.0\textwidth]{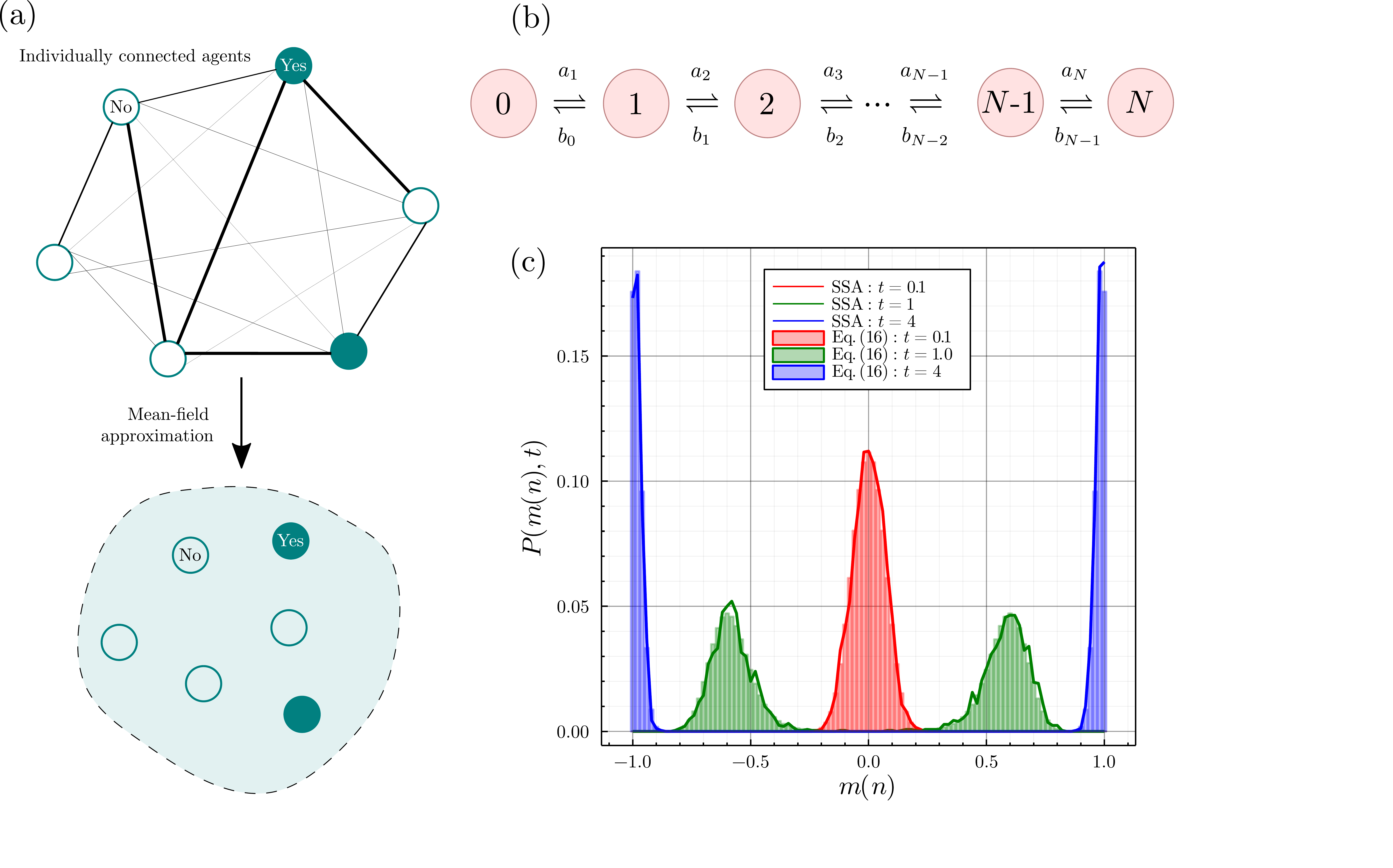}
\caption{(a) Illustration showing the effect of mean-field theory on a network of interconnected agents that influence each others decisions. In the mean-field case each agent is influenced by the same information $I(n,t)$ which they all contribute towards. (b) Diagram showing the 1D process that is equivalent to the mean-field case of the binary decision model. The number in each circle denotes the number of \textit{right deciding} agents, and the expressions above and below the arrows denote the associated probabilities of transitioning between each configuration. (c) Plots showing the correspondence between our analytic solution in Eq. (\ref{eq:exactsol1}) and Monte Carlo simulations provided by the SSA. The parameters for this time-evolution are $F=0,J=10,\alpha=0,\beta=1,\gamma = 1, N=100$ and $P(n,0)=\delta_{n,N/2}$. The probability distribution from the SSA is calculated from $2.5\times10^3$ trajectories. As $t\to\infty$ we see the emergence of a single steady-state consisting of two equal modes of height $\approx0.5$ at $m(0)=-1$ and $m(N)=1$. In the limit $N\to\infty$ this steady-state bimodality corresponds to symmetry breaking behaviour \cite{tauber2014critical}. Note that the SSA simulations in the solid lines show random perturbations due to the stochastic nature of the simulations, whereas the bars showing Eq.~\eqref{eq:exactsol1} do not show these deviations.}
\label{fig1}
\end{figure}

In the mean-field context one can interpret the meanings of $F(t)$, $J$ and $m(n)$ in many ways. One way, introduced in \cite{bouchaud2013crises}, is to interpret $F(t)$ as the cost of installing a new heating system and $J m(n(t))$ as the expected cost decrease as technical improvements cause more people to switch to the dominant technology. A similar interpretation will be used later on in Section \ref{sec:lock-ins} to discuss the phenomena of technology lock-ins. In many settings, mean-field approximations are reasonable for binary choice decisions: many decisions one makes on a day-to-day basis are influenced by many peers acting around an individual, for example, deciding whether to invest in cryptocurrency \cite{bouri2019herding}, or the language one uses in online reviews \cite{michael2014write}. As noted by Kirman \cite{kirman2010complex}, \textit{sociologists have long observed empirically that relational networks are likely to be much more connected than one might imagine}. However, in many cases network effects are very relevant, for example, children brought up in religious households are much more likely to be religious themselves \cite{cornwall1989determinants}, in which case the mean-field assumption is a poor one. We illustrate the concept of the mean-field assumption in Figure \ref{fig1}(a).

We now introduce the utility of agent $i$ as $U_i(n,t) \equiv S_i(t)I_i(\mathcal{S},t) = S_i(t)I(n,t)$, a quantity that a rational agent would want to maximise over any realisation of the model. We denote the situation wherein agents only care to maximise their own utility as a system of \textit{selfish agents} and generalise for more altruistic agents that are motivated by changes in the global utility in the following section. Now consider that the rate at which an agent changes their mind is dependent on \textit{the change in their own utility if they do change their mind}. For agent $i$, whose utility at time $t$ is $U(n,t)$, changing their decision from $S_i\to -S_i$ at time $t$ would result in the utility change, 
\begin{align}
    \Delta U_i(n,t) \equiv U_i(n-S_i,t) - U_i(n,t) = -2 S_i(t)I(n,t) + 2J/N. 
\end{align}
Note that $2J/N$ is a self-interaction term which comes from considering the change in $m(n)$ as $S_i\to -S_i$, a contribution that equally affects the rates of transition from $-1\to1$ and $1\to-1$, and that is negligible for $N\gg 1$. In the case where the change in the agent's decision is not assumed to affect $\Delta U_i(n,t)$ the term $2J/N$ can be ignored. If an agent had made decision $S_i(t) = -1$ (or $S_i(t)=1$) and the influence on them was $I(n,t)>0$ (or $I(n,t)<0$) then in changing their decision at $t$ they increase their utility; alternatively, if an agent had made decision $S_i(t) = -1$ (or $S_i(t)=1$) and the influence on them was $I(n,t)<0$ (or $I(n,t)>0$) then in changing their decision at $t$ they decrease their utility. Hence, a rational agent will (on average) make decisions such that their choice $S_i(t)$ agrees with the sign $I(n,t)$, at a rate (determined below) that is a monotonically increasing function of $\Delta U_i(n,t)$. \textit{More simply, the larger the change in the agent's utility $\Delta U_i(n,t)$ upon a decision change, the larger the rate at which an agent changes their decision.}

In this paper we choose decision rules based on detailed balance considerations \cite{tauber2014critical}. This means that we choose the rates with which agents update their decisions based on the assumption that as $t\to \infty$ the Boltzmann equilibrium distribution will be attained, i.e., $P_{eq}(S_i)\propto \exp(\beta U_i(n))$ where $\beta$ is the rationality of agent $i$ (discussed below). Enforcing detailed balance means the transition rates must satisfy,
\begin{align}\label{eq:transeqn}
    \frac{W_n(S_i(t)\to -S_i(t))}{W_{n-S_i}(-S_i(t)\to S_i(t))} = \frac{\exp(\beta U_i(n-S_i,t))}{\exp(\beta U_i(n,t))} = \exp(\beta \Delta U_i(n,t)),
\end{align}
where $W_n(S_i(t)\to -S_i(t))$ is the transition rate for agent $i$ to change their decision from $S_i\to -S_i$ at time $t$ given there are already $n$ agents deciding \textit{right} \cite{tauber2014critical}. That $W_n(S_i(t)\to -S_i(t))$ is a transition rate means that in a time interval $[t,t+\Delta t)$ an agent will change their decision with probability $W_n(S_i(t)\to -S_i(t))\Delta t$. There are several conventional ways to prescribe the transition rates from Eq.~\eqref{eq:transeqn}. A common way is to choose the Arrhenius form \cite{van1992stochastic} where the transition probabilities become $W_n(S_i(t)\to -S_i(t)) = \gamma \exp(\beta U_i(n-S_i,t))$ and $W_{n-S_i}(-S_i(t)\to S_i(t)) = \gamma \exp(\beta U_i(n,t))$, where $\gamma$ is a time scale parameter. An alternate form, and the form we use in this paper, are the Glauber/logit transition rates, which are given by \cite{glauber1963time,bouchaud2013crises,nadal1998formal}:
\begin{align}\label{eq:Glauber}
    W_n(S_i(t)\to -S_i(t)) = \frac{\gamma}{1+\exp(-\beta \Delta U_i(n,t))}
\end{align}
The choice of logit transition rates is motivated by \cite{nadal1998formal}, which derives Eq.~\eqref{eq:Glauber} using the maximum entropy principle. This assumes that agents make their choices based on a compromise between short term utility gain and a desire to sample other available choices. We note that the Arrhenius transition rates do not subscribe to this interpretation, since as is clear from their form, the transition rates are only dependent on the utility of the agent upon the change in decision; i.e., the agents do not consider the change in decision based on a knowledge of all decisions they could take, only the one they will take. In reality the proper form of the transition rates likely varies depending on the problem at hand. However, it seems likely that in the case of a binary decision an agent would need to first explore a range of decisions before they know the one that maximises their utility, hence we use Eq.~\eqref{eq:Glauber} for the rest of the paper. In other models the logit-like behaviour arises from other considerations, for example their presence in the \textit{future technology transformation} (FTT) models described in \cite{mercure2012ftt} is attributed to probabilistic nature of the cost of two competing technologies (see Fig.~1 of \cite{mercure2011global}). One also observes that the dependence of the propensity function $W_n$ on the change in utility is identical to that seen in classical choice theory \cite{anderson1992discrete}; this is by design, and the propensity of an agent to change their decision is proportional to the probability that the agent would indeed choose that option in choice theory.

Eq.~\eqref{eq:Glauber} allows us to explore the agent rationality $\beta$, which is a direct analogue of the inverse temperature commonly seen in equilibrium thermodynamics and statistical physics \cite{tauber2014critical}. As $\beta\to 0$ the agents become completely irrational and changes of decision occur at the same rate $\gamma/2$ regardless of an agent's utility change. Conversely, as $\beta\to \infty$ each agent is completely rational and Eq.~\eqref{eq:Glauber} becomes,
\begin{align}\nonumber
    \lim_{\beta\to\infty} W_n(S_i(t)\to -S_i(t)) = 
    \begin{cases} 
      0, & \Delta U_i(t)<0, \\
      \gamma, & \Delta U_i(t)>0, 
   \end{cases}
\end{align}
Hence, completely rational agents \textit{always} make decisions that agree with the influence upon them. For intermediate values of $\beta$ we obtain the typical `S-shaped' adoption curves with respect to $\Delta U_i(t)$, a common feature of many macroeconomic models including heterogeneous agents \cite{mercure2016modelling,mercure2012ftt,mccullen2013multiparameter}. We note that since $\beta$ is a constant we are implicitly assuming that each agent has the same level of rationality. Although in reality it may not be the case, for most situations we assume it to be a good approximation.

In the model described above `completely rational' agents correspond to agents that on their next change of decision, only choose the option that maximises their utility. So in a sense, even completely rational agents have limited foresight since they make decisions only based on the current state of the system, and not based on which decision in the long-run will optimise their utility---which would be the decision agreeing with the sign of $F$. Clearly, even these hyper-rational agents do not agree with the `perfect rationality' of agents observed in neoclassical economics (where agents even optimise for future conditions \cite{beinhocker2006origin}). The agents in the binary decision model more closely correspond to Keynesian agents with `fundamental uncertainty' in economies driven by short-term optimism (in our case short-term utility gain) \cite{king2015advanced}.

In contrast to the above, Kirman's model of ant rationality \cite{kirman1993ants} assumes a different form of propensities than the logit form seen in Eq.~\eqref{eq:Glauber}. Interpreting $S_i(t) =\pm 1$ now as two different sources of food and $n$ as the number of ants at the right-hand (+1) food source, ants switch food source due to two influences: the first is due to random switching with probability $\epsilon$ per unit time, the second due to recruitment by another ant with probability $\mu$ per unit time. The former is akin to an effective reaction with first-order kinetics, whereas the latter is akin to a reaction with second-order kinetics---i.e., the chance meeting of two ants is modelled as the probability of two molecules of the same species colliding. Following the laws of mass-action kinetics \cite{schnoerr2017approximation,gillespie2007stochastic,van1992stochastic}, the propensities for Kirman's ant model are then proportional to $W_n(-1\to+1)\propto (N-n)\epsilon+\frac{\mu (N-n)n}{N-1}$ and $W_n(+1\to-1)\propto n\epsilon+\frac{\mu (N-n)n}{N-1}$, as seen in \cite{moran2020schrodinger}. We will see below in Section \ref{sec:sol} that our method for solving the model of Brock and Durlauf in time also easily extends to Kirman's model, although we focus on the former model for the rest of the paper.


\subsection{Altruistic agents}\label{sec:altruistic}
We now generalise our mean-field model of economic agents such that they can take into account changes in the \textit{global utility} of all agents. We refer to this as \textit{agent altruism} since in this case agents do not only think about maximising their own utility but that of the entire population. This is an important quantity to take into account, and we do so in line with work conducted in \cite{grauwin2009competition}. In the context of binary decisions, having agents with altruistic attributes is somewhat realistic. Take the example we made above relating $F(t)$ to the cost of a heating system and $Jm(n)$ as the reduction in cost given more agents are using the technology. In this example altruistic agents could further benefit the utility of all individuals by more quickly arriving at a state where all agents are in agreement with one another. We define the global utility simply as the sum of the utilities of all agents,
\begin{align}\label{eq:globalutil}
    A(n,t) = \sum_{i=1}^N U_i(n,t).
\end{align}
Due to the mean-field interactions between the agents, a change in decision in one of the agents will change the utility of the individual agent by a different amount than the global utility. For example, say we have $\mathcal{S}_1=\{S_1,\dotsc,S_j,\dotsc,S_N\}$ and at time $t$ the system evolves to $\mathcal{S}_2=\{S_1,\dotsc,-S_j,\dotsc,S_N\}$: in general we find that $A(n-S_j,t)-A(n,t)\neq U_j(n-S_j,t)-U_j(n,t)$, unless there are no interactions between the agents. We consider $A(n-S_i,t)-A(n,t)$ explicitly below.

Now consider a system of entirely altruistic agents, whose decisions do not depend on their individual utilities but only on $A(t)$. As we did before for the selfish agents, we want to calculate the change in the $A(t)$ if agent $i$ changes their decision from $S_i \to -S_i$, i.e., $\Delta A_i(n,t) \equiv A_i(n-S_i,t) - A_i(n,t)$. Using Eq.~\eqref{eq:globalutil}, one finds that this is given by,
\begin{align}\label{eq:globaldiff}
    \Delta A_i(n,t) = \Delta U_i(n,t)-2J S_i \left(m(n) - \frac{S_i}{N}\right).
\end{align}
Hence, the change of the global utility is equal to the change in the individual utility of the agent minus a non-negligible term that accounts for the change in the utility of the rest of the population given agent $i$ has changed their decision. Note that the $m(n)-S_i/N$ term comes from considering the sum over all agents aside from agent $i$, $\sum_{j\neq i}S_j = N m(n) - S_i$, where the $S_i/N$ term is negligible for $N\gg1$. The extra term in Eq.~\eqref{eq:globaldiff} has an intuitive meaning, whose presence is explained by two mechanisms: (i) if $S_i$ (the original choice made by the agent) has the \textit{same} sign as $m(n)$ then $\Delta A_i < \Delta U_i$ since the agent has decided to go against the average opinion of all other agents encoded in $m(n)$; (ii) if $S_i$ has the \textit{opposite} sign to $m(n)$ then $\Delta A_i > \Delta U_i$ since the agent has decided to go with the average opinion, having previously been against it.

Following previous work \cite{grauwin2009competition} we now introduce the \textit{gain}, which gives a generalised form of utility change for agents that can be either entirely selfish, altruistic, or else somewhere in between. The gain for agent $i$ in changing their decision from $S_i\to-S_i$ is defined by,
\begin{align}\label{eq:gain}
    \mathcal{G}_i(n,t) &\equiv \Delta U_i (n,t)+ \alpha(\Delta A_i (n,t)- \Delta U_i(n,t))\\\nonumber
    &= -2S_i(F(t) + J m(n)(1+\alpha))+\frac{2(1+\alpha)J}{N},
\end{align}
where $\alpha$ is a parameter such that $\alpha = 0$ corresponds to a system of selfish agents, $\alpha = 1$ a system of altruistic agents, and $0<\alpha<1$ somewhere in between. We can finally write our most general transition rates including the effects of agent altruism,
\begin{align}\label{eq:genrule}
    W_n(S_i(t)\to -S_i(t)) = \frac{\gamma}{1+\exp(-\beta \mathcal{G}_i(n,t))},
\end{align}
which is essentially identical to the decision rule of Eq.~(1) in \cite{grauwin2009competition}. Curiously, and possibly somewhat intuitively, including the effects of altruism in our mean-field description acts to scale the agent-to-agent interaction strength by $1+\alpha$. As we explore in Section \ref{sec:metastab}, this scaling of the endogenous strength can have qualitative as well as quantitative effects on the model. It seems the mean-field approximation is responsible for the simplicity of this scaling, since in the general case of the model introduced in Section \ref{sec:econ_agents} the difference in utility change between altruistic agents and selfish ones is found to be the less simplistic and given by $\Delta A_i - \Delta U_i = -2\sum_{j\neq i}J_{ji}S_i S_j$. 
Note that for the case of generalised mean-field altruistic agents one can define a \textit{Hamiltonian} $\mathcal{H}(\mathcal{S})$, a function of the state of the system $\mathcal{S}$ such that, 
\begin{align}
    \Delta\mathcal{H}_i(\mathcal{S},t) \equiv \mathcal{H}(\{S_1,...,-S_i,...,S_N\},t) - \mathcal{H}(\{S_1,...,S_i,...,S_N\},t) = \mathcal{G}_i(n,t).
\end{align}
One finds this function is given by,
\begin{align}
    \mathcal{H}(\mathcal{S},t) = N m(n) \left(F(t) + \frac{1}{2}(\alpha+1)J m(n)\right),
\end{align}
where we remind the reader that $m(n) = (2n-N)/N = \sum_i S_i/N$. The existence of $\mathcal{H}(\mathcal{S})$ ensures that decision rules based on Eq.~\eqref{eq:genrule} satisfy detailed balance and that as $t\to\infty$ the equilibrium probability of each state is $P_{eq}(\mathcal{S})\propto \exp(\beta \mathcal{H}(\mathcal{S}))$ for all $0\leq \alpha\leq 1$.

\subsection{Master equation and analytical time-dependent solution}\label{sec:sol}
We can now proceed to solve the mean-field model analytically. First we map the model to a stochastic birth-death reaction scheme given by,
\begin{align}\label{eq:rs}
    L \xrightleftharpoons[n l(n)]{(N-n) r(n)} R,
\end{align}
where $r(n)$ and $l(n)$ are the transition rates at which each agent changes their decision from \textit{left to right} and \textit{right to left} respectively given there are $n$ right voting agents, and $L$ and $R$ are symbols representing agents who have decided left and right respectively. In this context one can state that a decision change of an agent is akin to a gain/loss of particles, where each particle denotes a right deciding agent \cite{van1992stochastic}. \textit{Importantly, from this birth-death mapping, where $F(t)\to F$ is a time-independent constant, it is clear that in the mean-field limit the number of right deciding agents $n$ completely determines the state of the system.} Hence, for the moment we set $F(t)\to F$, and discuss the case of time-dependent $F(t)$ later on. The rates $r(n)$ and $l(n)$ are defined by the decision rules we derived above in Eq.~\eqref{eq:genrule} and are explicitly given by,
\begin{align}\nonumber
    r(n) = W_n(-1 \to 1),\\
    l(n) = W_n(1 \to -1).
\end{align}
Note that the choice of the overall propensities $(N-n) r(n)$ and $n l(n)$ comes from mass-action kinetics: in short, the rate of gaining a new right deciding agent is proportional to the number $N-n$ of left deciding agents in the system \cite{van1992stochastic,schnoerr2017approximation}.

One can easily check that by writing the rate equation for the evolution of Eq.~\eqref{eq:rs}, in the large $N$ limit, as $t\to\infty$ one recovers the classic mean-field result of $m = \tanh(\beta(F + J(\alpha+1) m))$. However, we can do better than the rate equation describing the steady-state value of $m$, and can in fact solve for the probability of having $n$ right deciding agents at a time $t$, $\mathcal{P}(n,t)$. We first write the master equation corresponding to the process in \eqref{eq:rs},
\begin{align}\nonumber
    \partial_t \mathcal{P}(n,t) = [(N-(n-1))r(n-1)]\mathcal{P}(n-1,t)&+[(n+1)l(n+1)]\mathcal{P}(n +1,t)\\\label{eq:me}
    &-[(N-n)r(n)+nl(n)]\mathcal{P}(n,t).
\end{align} 
The master equation, otherwise known as the Kolmogorov forward equation, is a set of first-order differential equations describing the time evolution of being in discrete state $n$ at time $t$. Eq.~\eqref{eq:me} is a 1D master equation since it describing the evolution of only a single stochastic variable $n$. We outline the derivation of a 1D master equation from first principles in Appendix \ref{sec:mastereqn}. At steady-state, where $\partial_t \mathcal{P}(n,t\to\infty)=0$, 1D master equations are very well understood, and their properties are discussed in many textbooks \cite{van1992stochastic,gardiner2009stochastic,tauber2014critical}. Solving them in time is a much trickier problem. However, it is possible to solve Eq.~\eqref{eq:me} and we do so using the non-standard and useful method of \cite{smith2015general}. For readers that are less mathematically inclined we suggest that not too much time is spent on the derivation below; \textit{it is simply most important to realise that the mean-field binary decision model can be solved in time, and that its analytical solution is much quicker to compute than simulation based methods.} For consistency with \cite{smith2015general} publication we introduce the following notation: $a_n = (N-(n-1))r(n-1)$ and $b_n = (n+1)l(n+1)$, and we show the transitions between the microstates of the model using these propensities in Figure \ref{fig1}(b). Then, Eq.~(\ref{eq:me}) can be re-written as $\partial_t \vec{\mathcal{P}}(t) = \mathbf{A}\cdot \vec{\mathcal{P}}$, where $\vec{\mathcal{P}}(t)$ is the column vector $(\mathcal{P}(0,t),\mathcal{P}(1,t),...,\mathcal{P}(N,t))$ and $\mathbf{A}$ is the \textit{master operator}, a $(N+1)\times(N+1)$ tridiagonal matrix, given by,
\renewcommand*{\arraystretch}{1.7}
\begin{align}\label{eq:TRM}
    \mathbf{A} = 
    \begin{pmatrix}
        -a_1 & b_0 & & \\
        a_1 & -(b_0+a_2) & b_1&\\
            & a_2 & -(b_1+a_3)&\small{\ddots}\\
        & & \small{\ddots}& \small{\ddots}
    \end{pmatrix}.
\end{align}
We denote the eigenvalues of the matrix $\mathbf{A}$ as $\lambda_i$ for $i=1,2,...,N+1$, and from Perron-Frobenius theorem can state that the largest eigenvalue is $\lambda_1 = 0$ with $\lambda_1> \text{Re}(\lambda_2)\geq...\geq\text{Re}(\lambda_{N+1})$ \cite{pillai2005perron}. Note that the eigenvector corresponding to $\lambda_1 = 0$ is the steady-state probability distribution. Following some complex analysis on the resolvent of the master operator and judicious use of Cauchy's residue theorem in \cite{smith2015general} one then arrives at the solution of Eq.~(\ref{eq:me}) from an initial condition $\mathcal{P}(n,0) = \delta_{n,n_0}$, explicitly,
\begin{align}\label{eq:exactsol1}
    P(m(n),t|m(n_0),0)= \mathcal{P}(n,t|n_0,0) = 
    \begin{cases} 
      b_n\dotsm b_{n_0-1}\sum_{i=1}^{N+1}e^{\lambda_i t} \frac{p_n(\lambda_i)q_{n_0+2}(\lambda_i)}{\prod_{j\neq i} (\lambda_i-\lambda_j)}, & n<n_0, \\
      \sum_{i=1}^{N+1}e^{\lambda_i t} \frac{p_n(\lambda_i)q_{n_0+2}(\lambda_i)}{\prod_{j\neq i} (\lambda_i-\lambda_j)}, & n = n_0, \\
      a_{n_0+1}\dotsm a_n\sum_{i=1}^{N+1}e^{\lambda_i t} \frac{p_{n_0}(\lambda_i)q_{n+2}(\lambda_i)}{\prod_{j\neq i} (\lambda_i-\lambda_j)}, & n>n_0,
   \end{cases}
\end{align}
where the orthogonal polynomials $p_n$ and $q_n$ are recursively defined via,
\begin{align}
    p_1(y) &= 1,\; p_2(y) = y+a_1,\\
    p_i(y) &= (y+a_i+b_{i-2})p_{i-1}(y)-b_{i-2}a_{i-1}p_{i-2}(y),\\
    q_{N+2}(y) &= 1,\; q_{N+1} = y+b_{N-1},\\\label{eq:exactsolend}
    q_i(y) &= (y+a_i+b_{i-2})q_{i+1}(y) - b_{i-1}a_iq_{i+2}(y).
\end{align}
Eqs.~(\ref{eq:exactsol1}-\ref{eq:exactsolend}) constitute the analytical time-dependent solution for the mean-field Ising-Weidlich binary decision model with variably selfish-altruistic agents \cite{weidlich1971statistical}. For details regarding the derivation of this result we refer the reader to \cite{smith2015general}. In the most general case of usage of the model one needs to determine the eigenvalues of $\mathbf{A}$ computationally, which we do using the \code{eigvals} function in the Julia package \code{LinearAlgebra} \cite{bezanson2017julia}. However, since the eigenvectors are already implicit in the form of Eq.~\eqref{eq:exactsol1} we do not need to evaluate these computationally, and hence the analytical solution we utilise can be computed much more quickly than a finite state projection approach requiring matrix exponentiation \cite{munsky2006finite}. Note that the \code{eigvals} function that we use to compute the eigenvalues does not always return entirely real eigenvalues in every case (as one would physically expect) and often eigenvalues come in the form of a complex conjugate pairs. This is a common computational error, since we know theoretically that the eigenvalues should be real, and the resultant set of eigenvalues we obtain from \code{eigvals} is known as a \textit{pseudospectrum} \cite{iserles2019applications}, which arises since the eigenvalues of these matrices are very sensitive to small perturbations.

Importantly however, the \textit{usage of the pseudospectrum in Eqs.~(\ref{eq:exactsol1}-\ref{eq:exactsolend}) returns a normalised probability distribution that is indistinguishable from Monte Carlo simulation of the model}. The Monte Carlo simulation method we employ is known as the \textit{stochastic simulation algorithm} (SSA), and we detail it in Appendix \ref{sec:SSA}. Briefly, the SSA is a continuous time method often used in stochastic chemical kinetics to simulate chemical reactions. In our case, it outputs individual realisations of the master equation \eqref{eq:me}, which can then be replicated as an ensemble in order to construct the probability distribution, mean and variance of the process in time. In Figure \ref{fig1}(c) we show that our analytic time-dependent solution corresponds precisely to the distribution produced via the SSA for times near the initial condition through to near steady-state conditions. In Section \ref{sec:approximate_eigvals} we approximately calculate $\lambda_2$, the eigenvalue that determines the relaxation rate to the steady-state, using exact expressions for the switching times between lock-in states. Note that $\lambda_2^{-1}$ gives the time scale to reach the equilibrium. We further note that an analytical procedure to calculate the eigenvalues of our problem in a perturbative sense may be possible following methods used in \cite{tapias2020entropic}.

Eqs.~(\ref{eq:exactsol1}-\ref{eq:exactsolend}) assume that the initial condition is a fixed value of $n_0$ right deciding agents at $t=0$. However, often the initial state is not precisely known but is itself a distribution $\mathcal{Q}(n) \equiv \mathcal{P}(n,0)$. Using the laws of probability one determines the time evolution of $\mathcal{P}(n,t)$ with initial condition $\mathcal{Q}(n)$ as,
\begin{align}
    P(m(n),t|\mathbb{Q}(m(n)),0)= \mathcal{P}(n,t|\mathcal{Q}(n),0) = \sum_{n_0=0}^N \mathcal{Q}(n_0) \mathcal{P}(m(n),t|m(n_0),0),
\end{align}
where $\mathbb{Q}(m(n)) = \mathcal{Q}(n)$. In the common case where each agent is initially assigned a decision at random with probability $p_0$ the number of initially right deciding agents is drawn from a binomial distribution, i.e., $\mathcal{Q}(n) = \text{Bin}(n;N,p_0)$. The steady-state distribution reached as $t\to \infty$, i.e., $\mathcal{P}_s(n) = \mathcal{P}(n,t\to\infty)$, is widely known for birth-death processes with general rates and can be expressed as the following from Kirchhoff's theorem \cite{schnakenberg1976network},
\begin{align}\label{eq:exactSS}
    P_s(m(n))= \mathcal{P}_s(n) = \frac{(\prod_{i=1}^n a_i)(\prod_{i=n}^{N-1} b_i)}{\sum_j (\prod_{i=1}^j a_i)(\prod_{i=j}^{N-1} b_i)},
\end{align}
where we define the empty product $\prod_{j>i}^i$ as being equal to 1. $P_s(n)$ is independent of the initial condition, even for systems with long-time metastability/lock-in effects. Note that for the case of time-dependent $F(t)$ the solution above must be somewhat modified since $a_i$ and $b_i$ are no longer dependent only on $n$ but also explicitly on $t$. This case of having a zeitgeist that changes in time has important policy implications, and although it is not relevant for our work here we outline its solution in Appendix \ref{sec:TDZ}.

Finally, we note that if one interprets the economic agents as magnetic spins, the zeitgeist instead as a magnetic field, and $J$ as the mean-field exchange coupling (multiplied by the number of nearest neighbours), then our time-dependent solution above also provides the solution to the mean-field Ising model for a finite number of $N$ magnetic spins. This model of a magnet, famed for its simple equilibrium solution among other things, is still the subject of recent work \cite{ferrenberg2018pushing,morningstar2017deep,cervera2018exact}.

\section{Exploration of the model}\label{sec:metastab}
We can now use our analytic time-dependent solution to explore the model in different regimes of parameter space. Before delving into metastable behaviours, which will occupy us for the rest of this section, we detail two well known features of the model \cite{tauber2014critical,bouchaud2013crises,brock2001discrete}. The first is that, for $F=0$, the model exhibits a phase transition at a critical value of agent rationality $\beta_c = 1/J(1+\alpha)$. The origin of this critical value is explored in Appendix \ref{sec:criticality}, but essentially for $\beta>\beta_c$ one finds a symmetry breaking behaviour wherein agents either mostly become left or right deciding at steady-state and $P_s(n)$ is bimodal; whereas for $\beta\leq\beta_c$ the steady-state distribution $P_s(n)$ is monomodal with a peak at $m = 0$. At $\beta=\beta_c$ we see the characteristic shape of a flat-topped steady-state distribution centred at $m=0$ such that any small deviation above $\beta_c$ causes bimodality. Secondly, for $F\neq 0$ one finds that although there may still exist two equilibrium modes of behaviour deterministically, in reality the decision of same sign to $F$ is exponentially more favourable as $t\to\infty$ (shown explicitly in Section \ref{sec:approximate_eigvals}). Economically, what this means is that agents coalesce on the same decision when they are sufficiently rational and when collective effects are deemed at least as important as exogenous effects. A recent review of path-dependency effects similar to lock-ins is given in \cite{horst2008ergodicity}.

\subsection{Lock-ins}\label{sec:lock-ins}
Let's explore a new narrative of the model and suppose the two decisions left and right correspond to two competing technologies. In this scenario we can interpret the zeitgeist and interaction effects as follows. $F$ accounts for any exogenous effects on the agents, for example, the cost difference between the two technologies or the effects of changing social norms. Denoting $C_L$ and $C_R$ as the weight of these real and social costs of the left and right technologies, one can the break down $F$ into its constituent parts, i.e., $F = C_L-C_R$. On the other hand, the interaction term $J$ takes into account any endogenous changes arising from within the dynamics of the model, in particular: (1) learning effects as technology gets improved due to increased uptake \cite{nakicenovic1996freeing}, (2) any endogenous changes in price due to the increased uptake \cite{farmer2016predictable, nakicenovic1996freeing}, and (3) establishment effects, accounting for the ease of investment regardless of the price since the infrastructure needed for the technology is already there \cite{nakicenovic1996freeing}. If one can make sense of these different influences on $J$ empirically, then we can then break down $J$ into its constituent parts, $J = J_1+J_2+J_3$.

Importantly, the foresight of the agents is limited only to how their decision change at the current time changes their own (or global) utility, and they do not know \textit{a priori} which technology is the optimal one (i.e., they cannot distinguish between exogenous and endogenous influences). The state of maximal utility for the agents will clearly be that where all agents decide on the optimal technology (leading to the highest utility amongst all agents), which has the best combined trade-off between price and social appropriateness. However, it has been observed that this does not always happen, and path-dependency can lead to people making collective decisions that differ from the optimal one. This leads to so called \textit{lock-in} effects \cite{foxon2002technological}, a type of positive reinforcement metastable phenomenon. As highlighted by \cite{foxon2002technological}, several types of lock-in are observed in the real-world, including technological, institutional and carbon lock-ins. Clearly, understanding lock-in effects is important for policy considerations. For example, where there are competing technologies with some being more carbon efficient than others, consumers will be more likely to invest in the cheapest most common technology with the most infrastructure than the one that is better for the environment \cite{mercure2012ftt}. Coming up with strategies for how governments avoid such lock-ins is hence of great interest.

In some regions of parameter space lock-ins occur in the model of Brock and Durlauf, where there are 2 possible lock-in states $m^*$: the optimal one where $\text{sign}(m^*)=\text{sign}(F)$, and the non-optimal one where $\text{sign}(m^*)=-\text{sign}(F)$, \textit{which occur since the agent-to-agent interaction terms become dominant over the zeitgeist, i.e., switching to the alternate technology punishes the utility of an agent due to the collective effects}. However, whether a lock-in is observed or not depends on several factors. First, one must be in the regime wherein agents are rational enough to begin coalescing on either of the two choices, i.e., $\beta>\beta_c$. In this sense, $\beta_c$ is not only a property of the steady-state but also a requirement for certain types of transient phenomena. Second, $F$ should not be of too great a magnitude such that only 1 equilibrium solution of $m$ is present (see Appendix \ref{sec:criticality}). Thirdly, the initial condition should not be entirely weighted towards the steady-state distribution one finds as $t\to\infty$, i.e., since lock-ins are a path-dependent phenomenon there must be a viable opportunity for them to occur. 

In many complex systems metastability occurs, and it requires a time-dependent analysis of the problem at hand, since since considering only the steady-state is no longer sufficient. In many cases this makes analytical treatment very difficult since time-dependent problems are rarely soluble. However, in this case we do have a solution, and in the sections that follow we proceed analytically. In the next section we derive the relaxation time scales to the steady-state for systems exhibiting lock-ins by determining the switching times between the two lock-in states. In the following section we will often refer to decisions as technologies, in line with the technology application drawn above.

\subsection{Escape from non-optimal decisions}\label{sec:approximate_eigvals}
\subsubsection{Relating the relaxation time scale to the first passage time}
\begin{figure}[h!]
\captionsetup{width=1.0\textwidth}
\centering
\includegraphics[width=0.8\textwidth]{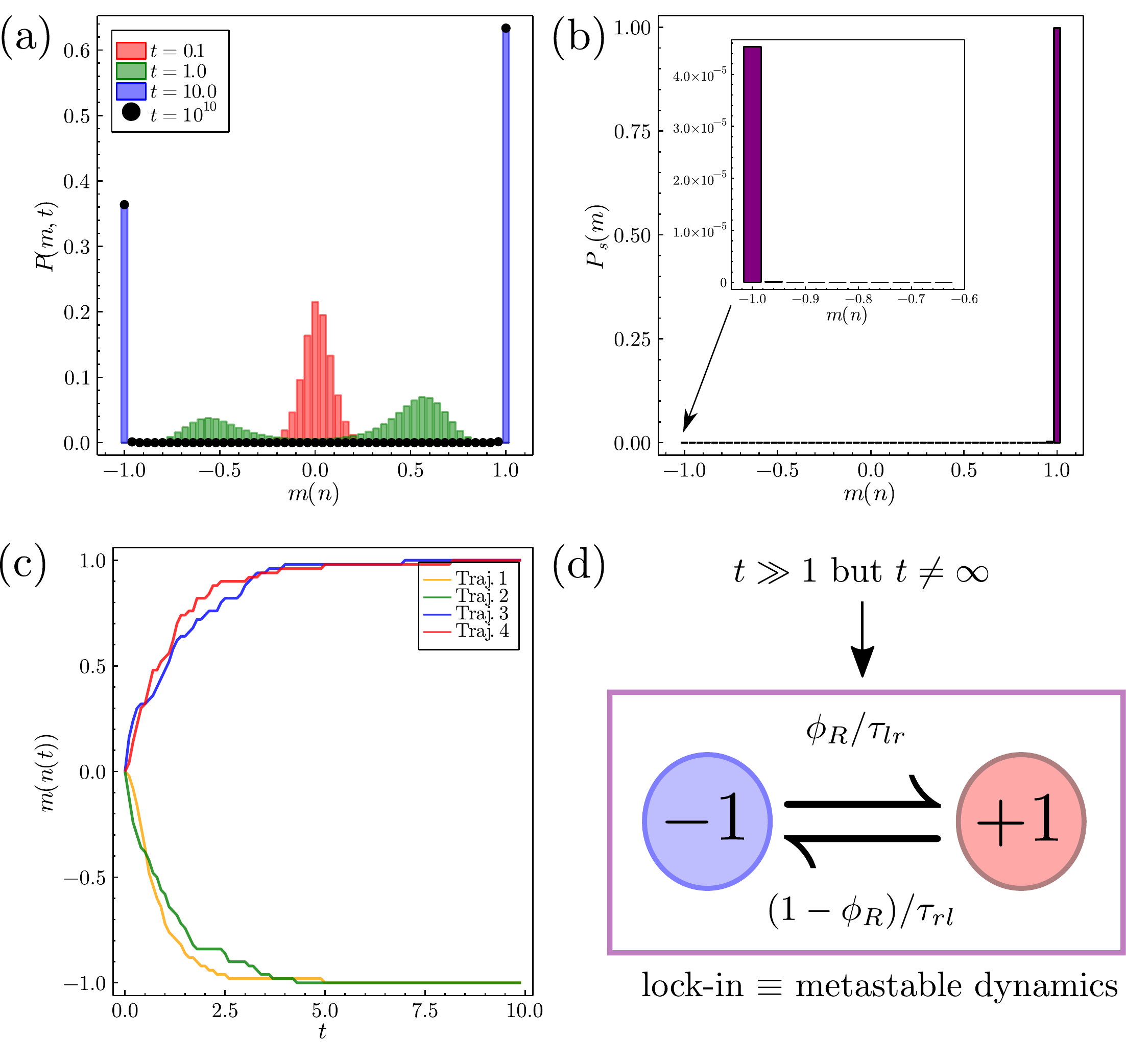}
\caption{Plots showing the lock-in phenomena in the mean-field model. (a) For $F>0$ and $\beta>\beta_c$ we show the evolution of the probability distribution given by Eq.~\eqref{eq:exactsol1} from an initial condition at $n=N/2$. Beyond the initial condition we see the emergence of two modes of behaviour, the agents either increasingly choose the left or the right technology. The evolution of the probability distribution becomes very slow for $t>10$ and the distribution at $t=10^{10}$ is indistinguishable from that at $t=10$. (b) A plot of the analytic steady-state distribution from Eq.~\eqref{eq:exactSS} shows that in the true steady-state limit almost all the agents will be locked into the right technology. Importantly, this is completely distinct from the time-dependent solution even at large times. Note the presence of a small mode at the unfavoured left-hand technology seen in the inset. (c) Individual SSA trajectories for the system showing the lock-in effect on individual populations. (d) As $t\gg 1$ the dynamics of the system becomes metastable and can be approximately mapped to a two state process with very small transition rates. Parameters for plots in this figure are $F=0.1,J=5,\alpha=0,\beta=1,\gamma = 1 \text{ and } N=50$.}
\label{fig2}
\end{figure}

We now look at the lock-in mechanism in more detail using our analytical solution. The general case of metastability in bistable systems is covered by van Kampen \cite{van1992stochastic}, who does so in three steps. These steps are vital to the explanation of the lock-in mechanism, and in determining the relaxation time scale $\lambda_2^{-1}$ to the steady-state as $t\to\infty$. Step 1 describes the evolution of the initial distribution. For $t\gtrsim0$, the distribution broadens and probability modes have not yet developed at the equilibria. This is seen in Fig.~\ref{fig2}(a) for the distribution in red at $t=0.1$ (the initial condition at $t=0$ being that of equal numbers of agents deciding left and right). Step 2 is that two distinct modes have developed in $P(m(n),t)$, which decompose into left and right parts respectively. How the distribution has decomposed is dependent on the initial condition and can be seen in Fig.~\ref{fig2}(a) for the distribution in green. Finally for $t\gg 1$, step 3 is that each mode has developed its \textit{local equilibrium shape} and that the transfer of probability between the modes becomes very small. This is shown in Fig.~\ref{fig2}(a) for the blue distribution and for the black dots, showing that the distributions between $t=10$ and $t=10^{10}$ are indistinguishable. At times $t\gg \lambda_3^{-1}$, where $\lambda_3$ is the third smallest eigenvalue of the master operator in Eq.~\eqref{eq:TRM}, the process can then be effectively modelled as a two-state process between the equilibria, with very small rates of probability transfer between the modes, as illustrated in Fig.~\ref{fig2}(d). In the following our main interest is in finding these rates of transfer between the behavioural modes and the overall rate of relaxation towards the steady-state.

Before proceeding, we illustrate how to calculate the equilibria values of $m$, including the unstable one, via the steady-state distribution alone (see Fig.~\ref{fig2}(b)). For $\beta>\beta_c$, the stable equilibria $m_-$ and $m_+$ are the maxima of the steady-state distribution ($-1$ and $1$ in Fig.~\ref{fig2}(b)) with the larger of the two modes being the optimal equilibrium value. The intermediate unstable equilibrium point $m_u$ is the minima of the distribution (found at $N/2$ for Fig.~\ref{fig2}(b)), and a system found either side of $m_u$ will likely drift towards the corresponding stable equilibrium point. We will denote the critical \textit{number} of right deciding agents as $n_u = N(m_u+1)/2$. Note that for $\beta\leq\beta_c$ (see Appendix \ref{sec:criticality}), or for $|F|> J(\alpha+1)$ in the $\beta\gg 1$ limit (see Appendix \ref{sec:bouchaud_calc}), there is only one equilibrium value of $m$.

Defining $\pi_L(t)=\sum_{m<m_u}P(m,t)$ and $\pi_R(t)=\sum_{m>m_u}P(m,t)$ as the probabilities to be in the left and right modes respectively, following the approximative two-state process in Fig.~\ref{fig2}(d), we can then assert that,
\begin{align}\label{eq:metastab_eq}
    \partial_t \pi_L(t) = -\partial_t \pi_R(t) = -\frac{\phi_R\pi_L(t) }{\tau_{lr}}+\frac{(1-\phi_R)\pi_R(t) }{\tau_{rl}},
\end{align}
where $\tau_{lr}$ and $\tau_{rl}$ are the mean times to escape the from equilibria at the left and right modes respectively (i.e., starting at the equilibria, how long it takes to reach $m_u$), and $\phi_R$ is the probability that if the system starts at $m(0)=m_u$ that the agents initially coalesce on the right technology. We derive $\phi_R$ in Appendix \ref{sec:FP} (using similar methods as the calculation of the mean first passage time in the next section). This probability enters the rates of transition to the other mode since $\tau_{lr}$ and $\tau_{rl}$ only give the times to escape from each mode, which must be scaled by the probability to enter the other mode, since upon reaching $m_u$ the agents could also revert back to the decision they inhabited before. Eventually, $\pi_L$ and $\pi_R$ reach their stationary values which are determined by $\partial_t \pi^s_L = 0$, explicitly, 
\begin{align}
    \frac{\pi^s_L}{\pi^s_R} = \frac{\tau_{lr}(1-\phi_R)}{\tau_{rl}\phi_R}.
\end{align}
This equation helps us interpret the results of Fig.~\ref{fig2}(b): for the chosen parameter set $\tau_{lr}\ll\tau_{rl}$: once the agents have made the optimal decision they will not change it on any reasonable time scale. Note that even at the steady-state there is a mode at $m=-1$ (see inset Fig.~\ref{fig2}(b)), and that the \textit{shape} of the left-hand distribution will be near identical to the shape of the distribution at $t=10$ in Fig.~ \ref{fig2}(a), but is of much smaller magnitude. The three steps described by van Kampen can also be seen on a SSA trajectory level in Fig.~\ref{fig2}(c): after an initial period of indecisiveness the individual population trajectories eventually coalesce on either the optimal technology at $m=1$ (with slightly greater probability) or else the sub-optimal technology at $m=-1$.

Before calculating the switching times between the equilibria, one can ask how these switching times are related to the rate of relaxation to the steady-state. Namely, how do $\tau_{lr}$ and $\tau_{rl}$ relate to the smallest magnitude non-zero eigenvalue $\lambda_2$? We know that for $t\gg \lambda_3^{-1}$ the only relevant time scale is the relaxation time scale from the metastable state to the steady-state, hence we can write,
\begin{align}\label{eq:p_ass}
    P(m,t)\sim P_s(m) + \exp(-\lambda_2 t)\Phi_2(m),
\end{align}
where $\Phi_2(m)$ is the eigenvector of $\mathbf{A}$ (see Eq.~\eqref{eq:TRM}) corresponding to the eigenvalue $\lambda_2$. Inserting Eq.~\eqref{eq:p_ass} into Eq.~\eqref{eq:metastab_eq} we find that,
\begin{align}\label{eq:eigenvalue2}
    \lambda_2 \sim \frac{\phi_R}{\tau_{lr}}+\frac{(1-\phi_R)}{\tau_{rl}},
\end{align}
where we have used the fact that $\sum_m \Phi_2(m) = 0$, which comes from summing Eq.~\eqref{eq:p_ass} over all $m$ and enforcing normalisation conditions on $P(m,t)$ and $P_s(m)$. As Bouchaud showed in \cite{bouchaud2013crises}, one can approximate the master equation in \eqref{eq:me} as a Fokker-Planck equation \cite{gardiner2009stochastic,van1992stochastic}, hence finding that the mean first passage times are proportional to
\begin{align}\label{eq:approx_tau1}
    \tau_{lr} &\propto \exp\left(N(1-F/J(1+\alpha))\right),\\\label{eq:approx_tau2}
    \tau_{rl} &\propto \exp\left(N(1+F/J(1+\alpha))\right),
\end{align}
whose full derivation we show in Appendix \ref{sec:bouchaud_calc}. The utility of this result is its ease of interpretation: where there are two stable equilibria the mean times for switching between them are \textit{exponential in the number of agents}. 

\subsubsection{A better approximation of the relaxation time scale}\label{sec:exactdet}
\begin{figure}[h!]
\captionsetup{width=1.0\textwidth}
\centering
\includegraphics[width=0.8\textwidth]{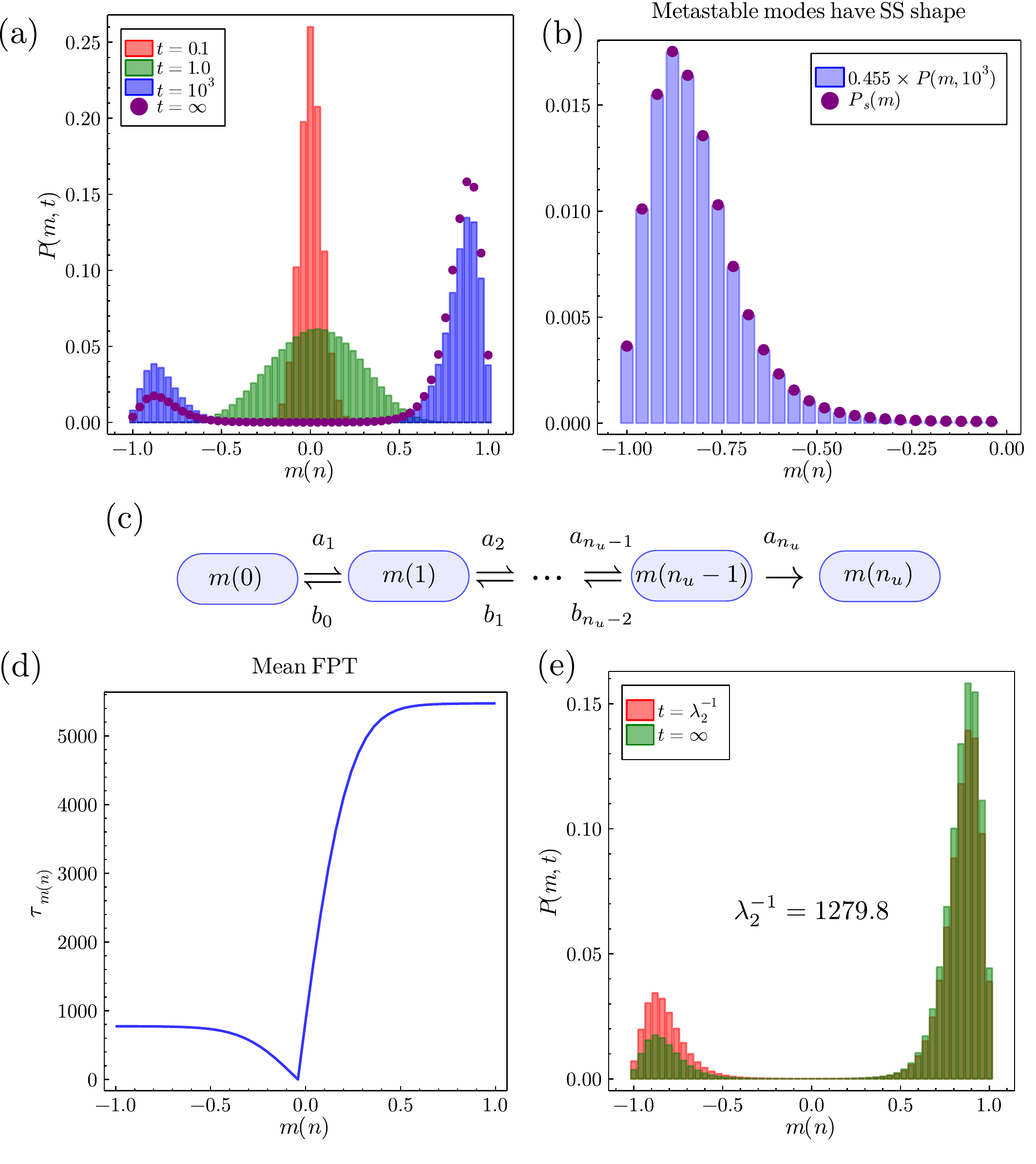}
\caption{Figure showing the coalescence of the agents onto stable equilibria where the modes are not found at $m=\pm 1$ and fluctuations are present, for parameters $F=0.025,J=1.5,\alpha=0,\beta=1,\gamma = 1 \text{ and } N=50$ using the analytic solution from Eq.~\eqref{eq:exactsol1}. (a) Plots of $P(m,t)$ at times from near the initial condition at $m = 0$ to the steady-state. Note that for $t\gtrsim 3\times10^3$ the time-dependent solution becomes indistinguishable from the steady-state. (b) Plot showing that the metastable modes have the same \textit{shape} as the modes of the steady-state distribution. We show that the right-hand mode rescaled by a prefactor ($= 0.455$ here for $t=10^3$) becomes indistinguishable from the steady-state distribution. (c) New dynamics considered for the calculation of the mean first passage time $\tau_{lr}$. Note that a similar, but separate, diagram can be drawn for the calculation of $\tau_{rl}$. (d) Plot of the mean first passage time to reach $m_u$ given one starts at $m(n)$. It is clear that since $F>0$ the mean first passage times to hit $m_u$ are greater for $m>0$. (e) Plot of the time-dependent distribution at $t=\lambda_2^{-1}= 1279.8$ (using our approximation) versus the steady-state distribution.}
\label{fig3}
\end{figure}

Eqs.~\eqref{eq:approx_tau1}-\eqref{eq:approx_tau2} are very good approximations in the $\beta\gg 1$ and $N\gg1$ limits where $m_{\pm} = \pm 1$. However, there are some cases where the distributions $\pi_L(t)$ and $\pi_R(t)$ have a greater variance and are less peaked (as illustrated in Fig.~\ref{fig3}(a)) in which case contributions to the time needed to pass the potential barrier at $m_u$ must be taken from multiple values of $m$. We first define the conditional normalised probability distributions of having $m$ inside each of the the metastable phases as $\rho_l(m<m_u) = P_s(m)/\sum_{m<m_u}P_s(m)$ and $\rho_r(m>m_u) = P_s(m)/\sum_{m>m_u}P_s(m)$ for the left and right equilibria respectively. We determine this from the intuition that the metastable probability modes are the same as those found in the steady-state distribution up to a multiplicative pre-factor, an intuition that we confirm in Fig.~\ref{fig3}(b). One can then define the mean times to switch between the two equilibrium modes at $m<m_u$ and $m>m_u$ as weighted sums over the conditional distributions,
\begin{align}\label{eq:taulr1}
    \tau_{lr} &= \sum_{n=0}^{n_u-1}\rho_l(m(n))\tau_n,\\\label{eq:taulr2}
    \tau_{rl} &= \sum_{n=n_u+1}^{N}\rho_r(m(n))\tau_n,
\end{align}
where $\tau_n$ is the mean first passage time to reach $n_u$ given one starts with $n$ agents deciding on the right-hand technology. In our case it is possible to find the values of $\tau_n$ \textit{exactly}, and we do so in line with methods shown in \cite{van1992stochastic,ashcroft2016metastable,antal2006fixation}.

Consider again the microscopic transitions previously seen in Fig.~\ref{fig1}(b). For this birth-death process one can write a \textit{backward equation}, which is formally the adjoint equation to the master equation, commonly used for first passage processes. For our purposes, it is more convenient to use the discrete time backward equation given by \cite{van1992stochastic,ashcroft2016metastable},
\begin{align}\label{eq:bme}
    Q_{j,i}(t+\Delta t) = a_{i+1} \Delta t Q_{j,i+1}(t) + b_{i-1} \Delta t Q_{j,i-1}(t) + (1-(a_{i+1}+b_{i-1})\Delta t)Q_{j,i}(t),
\end{align}
where $a_i = (N-(i-1))r(i-1)$, $b_i = (i+1)l(i+1)$, $Q_{j,i}(t)$ is the probability of being found with $j$ agents deciding on the right technology a period of time $t$ after being found with $i$ right technology deciding agents, and $\Delta t$ is the time step (which is taken to zero in the continuous time limit). Note that the absolute time $t_n$ at time step $n$ is defined by $t_n = n\Delta t$.

The mean first passage times for hitting $m_u$ from $m<m_u$ and $m>m_u$ must be considered separately. In the following we show the calculation for $m<m_u$, although the procedure is analogous for $m>m_u$. Consider only the states of the system $m(n)<m_u$, where we define $m_u = m(n_u)$ with $n_u$ being the number of agents deciding for the right technology at the unstable equilibrium point. We define a new Markov process by setting $b_{n_u-1} = 0$, which defines $m_u$ as an absorbing boundary \cite{van1992stochastic,braichenko2021distinguishing}. A schematic of these new dynamics is show in Fig.~\ref{fig3}(c). From the backward equation for this new process we then define $\psi_i(t) = Q_{n_u,i}(t)$ as the \textit{cumulative} probability of reaching $n_u$ a time $t$ after having $i$ right deciding agents. Hence, the probability of hitting $n_u$ \textit{at} time $t$ is given by $\psi_i(t)-\psi_i(t-\Delta t)$, and the mean first passage time $\tau_{n}$ is given by a weighted sum over these probabilities,
\begin{align}
    \tau_{n} = \sum_{i=0}^\infty i\Delta t(\psi_n(i\Delta t)-\psi_n((i-1)\Delta t)),
\end{align}
with $\psi_n(-\Delta t) = 0$. From the backward equation we find the recursive relationship for $\psi_i(t)$,
\begin{align}\label{eq:psi_eq}
    \psi_i(t+\Delta t) = a_{i+1} \Delta t \psi_{i+1}(t) + b_{i-1} \Delta t \psi_{i-1}(t) + (1-(a_{i+1}+b_{i-1})\Delta t)\psi_i(t),
\end{align}
which has the intuitive interpretation that the probability of reaching $m_u$ at time $t+\Delta t$ is the probability of hopping to $i+1$ and reaching $m_u$ from there, \textit{plus} the probability of hopping to $i-1$ and reaching $m_u$ from there, \textit{plus} the probability of not hopping and reaching $m_u$ from $i$ \cite{ashcroft2016metastable}. If we subtract $\psi_i(t)$ from both sides of Eq.~\eqref{eq:psi_eq} and sum over all $t$ we then get a recursion relation for the mean first passage time,
\begin{align}\label{eq:tauRR}
    a_{i+1} \tau_{i+1} + b_{i-1} \tau_{i-1} -(a_{i+1}+b_{i-1})\tau_{i} = -1.
\end{align}
One can perform similar calculations for the higher order moments of the first passage time distribution, although we do not consider them here (see \cite{ashcroft2016metastable} for more details). We have two boundary conditions on this recursion relation. The first is $\tau_{n_u} = 0$, which reflects the fact that if one starts at $n_u$ the time taken to reach it is obviously 0. The second is less obvious in that it comes from physical conditions at the left boundary and is $\tau_1 - \tau_0 = -1/a_1$. This reflects the fact that in the state $n=0$ there is only one possible move to $n=1$, which occurs with average time $1/a_1$. Now, we introduce the difference variable $\eta_i = \tau_i-\tau_{i-1}$ which transforms Eq.~\eqref{eq:tauRR} into,
\begin{align}
    \eta_i = \frac{b_{i-2}}{a_{i}}\eta_{i-1}-\frac{1}{a_{i}},
\end{align}
subject to $\eta_1 = -1/a_1$. This can be solved recursively and the result is,
\begin{align}
    \eta_i = -\sum_{j=1}^i\frac{1}{a_{j}}\prod_{k=j}^{i-1}\frac{b_{k-1}}{a_{k+1}},
\end{align}
where again we note that the empty product is equal to 1. Finally, to find the $\tau_n$ from $\eta_n$ we see that,
\begin{align}\label{eq:tau_less_nu}
    \tau_{n<n_u} = -\sum_{i=n+1}^{n_u}\eta_i = \sum_{i=n+1}^{n_u}\sum_{j=1}^i\frac{1}{a_{j}}\prod_{k=j}^{i-1}\frac{b_{k-1}}{a_{k+1}},
\end{align}
where we have used the fact that $\tau_{n_u} = 0$.. Using the same approach (or via symmetry considerations) one can derive the case of $n>n_u$, whose result is,
\begin{align}\label{eq:tau_gtr_nu}
    \tau_{N-n>n_u} = \sum_{i=n+1}^{N-n_u}\sum_{j=1}^i\frac{1}{b_{N-j}}\prod_{k=j}^{i-1}\frac{a_{N-(k-1)}}{b_{N-(k+1)}}.
\end{align}
The results derived in Eqs.~\eqref{eq:tau_less_nu} and \eqref{eq:tau_gtr_nu} are \textit{exact}. If one wishes for a simpler result, it is possible to simplify the expressions in the large $N$ limit by approximating the product term in Eqs.~\eqref{eq:tau_less_nu} and \eqref{eq:tau_gtr_nu} with $\prod_k \gamma_k(n) = \exp(\sum_k \log(\gamma_k(n)))\sim \exp(N\int\log(\gamma_k(n))dn)$, where we have explicitly included the dependence on $n$ for clarity (see \cite{antal2006fixation} for more details). Doing so gives similar exponential dependence on $N$ as we explored in Eqs.~\eqref{eq:approx_tau1}-\eqref{eq:approx_tau2}, and hence we do not show the result of the large $N$ approximations on Eqs.~\eqref{eq:tau_less_nu} and \eqref{eq:tau_gtr_nu} here. 

We show a plot of $\tau_{m(n)} \equiv \tau_n$ in Fig.~\ref{fig3}(d) for $F>0$, where we clearly see that it is more difficult to escape from the optimal choice technology than the sub-optimal since on average one has to wait much longer to escape the more optimal collective decision. The values of $\tau_n$ determined via Eqs.~\eqref{eq:tau_less_nu} and \eqref{eq:tau_gtr_nu} can then be used in combination with Eqs.~\eqref{eq:taulr1} and \eqref{eq:taulr2} in order to find the relaxation time scale $\lambda_2$ in Eq.~\eqref{eq:eigenvalue2}. Note that although this determination of $\lambda_2$ is still formally an approximation it is very accurate, and we show in Fig.~\ref{fig3}(e) that for the time-dependent solution at $t=\lambda_2^{-1}$ it results in a distribution very close to the steady-state distribution. Computationally, using \code{eigvals}, we find $\lambda_2^{-1}\sim 1288.8$ for Fig.~\ref{fig3}, whereas our approximation gives $\lambda_2^{-1}\sim 1279.8$, hence only slightly underestimates the relaxation time to reach the steady-state. Our approximation is an underestimation since the relaxation time since it does not take into account the initial relaxation into the metastable state of $\mathcal{O}(\lambda_3^{-1})$. 

Since for Fig.~\ref{fig3} we have $N=50$ agents, one can also conclude that our method works well \textit{even when not in the large $N$ limit}. Moreover, our approximation is even better for systems such as that seen in Fig.~\ref{fig2}, where the modes at $n_-$ and $n_+$ are very strongly peaked, since this type of system more closely corresponds to the approximations we have made in our calculation of $\lambda_2$ as the time taken to reach the metastable regime is even shorter compared to the time taken to reach the steady-state. The disparity seen between our approximation for $\lambda_2$ for the parameters in Fig.~\ref{fig3} constitutes a worst case scenario of the calculation of $\lambda_2$. Eq.~\eqref{eq:p_ass} then constitutes an approximate solution for the bimodal regime of the mean-field model that does not require calculation of the fast decaying eigenvalues computationally, valid where $t\gg \lambda_3^{-1}$. We finally observe that our calculation of $\lambda_2$ gives us an approximate time-dependent probability distribution, $P(m,t)\sim P_s(m) + \exp(-\lambda_2 t)\Phi_2(m)$, \textit{which does not require the calculation of any eigenvalues and is computationally valid in the metastable regime where all other eigenvalues of order $\mathcal{O}(\lambda_3^{-1})$ or less have decayed away.} Practically, in order to implement this result, upon realising that we have $\lambda_1 = 0$ and $\lambda_2$ we then substitute these into Eq.~\eqref{eq:exactsol1} and where $t\gg \lambda_3^{-1}$ we can safely ignore the contributions from $i\geq 3$ in the sum.

\section{Model calibration}\label{sec:cali}
Brock and Durlauf \cite{brock2001discrete} used their equilibrium solution for econometric analysis and model calibration. However, as we have emphasised in the previous section, an equilibrium solution is often not applicable on any realistic time scale, even for small numbers of agents, and hence time-dependent dynamics of the binary choice model must be considered. Generally speaking, finding an analytic time-dependent likelihood function for an agent-based model is a rarity \cite{sornette2014physics}, however our time-dependent solution for the probability distribution allows us to construct one for the mean-field binary decision model,\textit{ allowing us to conduct practical model calibration in a short time}. This makes our solution very useful to researchers who do not have access to vast amounts of computing power. We then test our calibration procedure on simulated data with known parameters in order to assess the conditions on the data necessary to provide reliable calibration.

\subsection{Construction of likelihood function}
We begin by defining the likelihood in the standard way. Say we have a set of $L$ data points $\mathcal{M}=\{m(t_i)\}$ for $i\in\{1,2,...,L\}$, measured at times $\{t_1,t_2,...,t_L\}$ describing the evolution of the order parameter $m(n(t))$ over some time period. In general one cannot assume this process is at equilibrium and the evolution from the initial condition at $m(t_1)$ does not follow a steady-state trajectory. This requires us to use Eq.~\eqref{eq:exactsol1} in order to calculate the \textit{probability of observing the trajectory $\mathcal{M}$ given some assumed parameter set $\theta = \{F, J,\gamma\}$}, known as the likelihood, and is given by,
\begin{align}
    L(\mathcal{M}|\theta) = \prod_{i=1}^{L}P_\theta(m(n(t_i)))
\end{align}
where $P_\theta(m(n(t_i)))$ is Eq.~\eqref{eq:exactsol1} evaluated for parameter set $\theta$ at time $t_i$. The likelihood function is expressed as a product of probabilities over the whole time series since it is the probability of observing $m(t_1)$ at $t_1$ \textit{and} $m(t_2)$ at $t_2$ and so on (hence from the laws of probability is multiplicative). Note that since $\beta$ pre-multiplies $F$ and $J$ and $(1+\alpha)$ pre-multiplies $\beta J$ in the utility function, $\beta$ or $\alpha$ cannot be inferred separately but only $\beta F$ and $\beta (1+\alpha) J$ can be inferred \cite{brock2001discrete}. Hence for calibration we set $\beta = 1$ and $\alpha = 0$ and infer only $F$, $J$ and $\gamma$. The aim of the calibration procedure is then to find the parameter set $\theta^\star$ that maximises $L(\mathcal{M}|\theta)$ with respect to $\theta$. Generally, since the likelihoods are generally very small quantities, it is more computationally convenient to instead minimise the negative log likelihood $-\ln(L(\mathcal{M}|\theta))$ which generally takes values $\gg1$. In most cases (including ours), the parameter set corresponding to the minimum value of $-\ln(L(\mathcal{M}|\theta))$ cannot be analytically determined and hence one must proceed algorithmically by testing many different values of $\theta$ via an optimisation algorithm. In this paper we utilise the \textit{adaptive differential evolution} optimiser from the Julia package \code{BlackBoxOptim} \cite{qin2005self,bbobest}, which is determined to be the best likelihood optimisation algorithm compared to several other state-of-the-art methods \cite{bbobest}. Using this algorithm, the optimal parameter set $\theta^\star$ is then determined through,
\begin{align}
    \theta^\star= \underset{{\mathbf{\theta} \in \Theta}}{ \arg \min} \left(-\log{L(\mathcal{M}|\theta)}\right)
\end{align}
where $\Theta$ is the set of all possible parameters in the optimisation range selected. In the calibration that follows the true parameter set for the simulated data is $\theta_{\text{true}} = \{ F=0.025,J=1.5,\gamma=1.0\}$ (same as in Fig.~\ref{fig3}) and the parameter ranges of optimisation for the set $\Theta$ were chosen such that $F\in [-2,2]$, $J\in[\exp(-2),\exp(2)]$ and $\gamma\in[\exp(-1),\exp(1)]$, where parameters $J$ and $\gamma$ are optimised in log-space. For the purpose of analysing the errors in the calibration procedure below we define,
\begin{align}\label{eq:toterr}
    E_{tot} &= \sum_i\left| \frac{\theta_{\text{true},i}-\theta^\star_i}{\theta_{\text{true},i}} \right|,\\\label{eq:ferr}
    f &= \left|\frac{(F_\text{true}/J_\text{true}) - (F^\star/J^\star)}{F_\text{true}/J_\text{true}}\right|, 
\end{align}
where $E_{tot}$ is the total error on the inference procedure and $f$ is the error in the fraction between $F$ and $J$. In most econometric analyses it will be the error in $f$ which is most important since it tells us whether one has properly inferred the magnitude of the influence of exogenous versus endogenous effects on the agents. These error functions will then allow us to determine the conditions necessary of data such that one can conduct a valid model calibration. We note that $\theta^\star$ does not generally correspond the global minimum of $-\ln(L(\mathcal{M}|\theta))$, but instead a \textit{good local minimum}. This is due to the complexity of the function $-\ln(L(\mathcal{M}|\theta))$, a problem for which there is no known solution to produce the global optimum in a reasonable computational time \cite{horst2000introduction}. The number of optimisation steps we choose for the adaptive differential algorithm is $500$, with a population size of $1000$ such that each calibration procedure took less than $1500s$ on a laptop with Intel Core i7-8650U CPU @ 1.90GHz × 8 running Ubuntu 18.04.5 LTS.

\subsection{Calibration from a single trajectory}
In many real-world situations data is only collected once for a particular event. For example, national elections do not have multiple realisations, hence if one was to collect the data of the percentage of votes for Democrats or Republicans at all elections from 1868 (from when only Democrats \textit{or} Republicans have won the presidential vote) then one ends up with a single trajectory of data. Therefore, it is important to assess the conditions under which calibration procedures are accurate with respect to data for which the parameters are known. For this purpose we utilise the SSA to provide stochastic trajectories for a particular realisation of our mean-field system for the parameter set explored in Fig.~\ref{fig3}, which notably expresses bimodality.

\begin{figure}[h!]
\captionsetup{width=1.0\textwidth}
\centering
\includegraphics[width=1.0\textwidth]{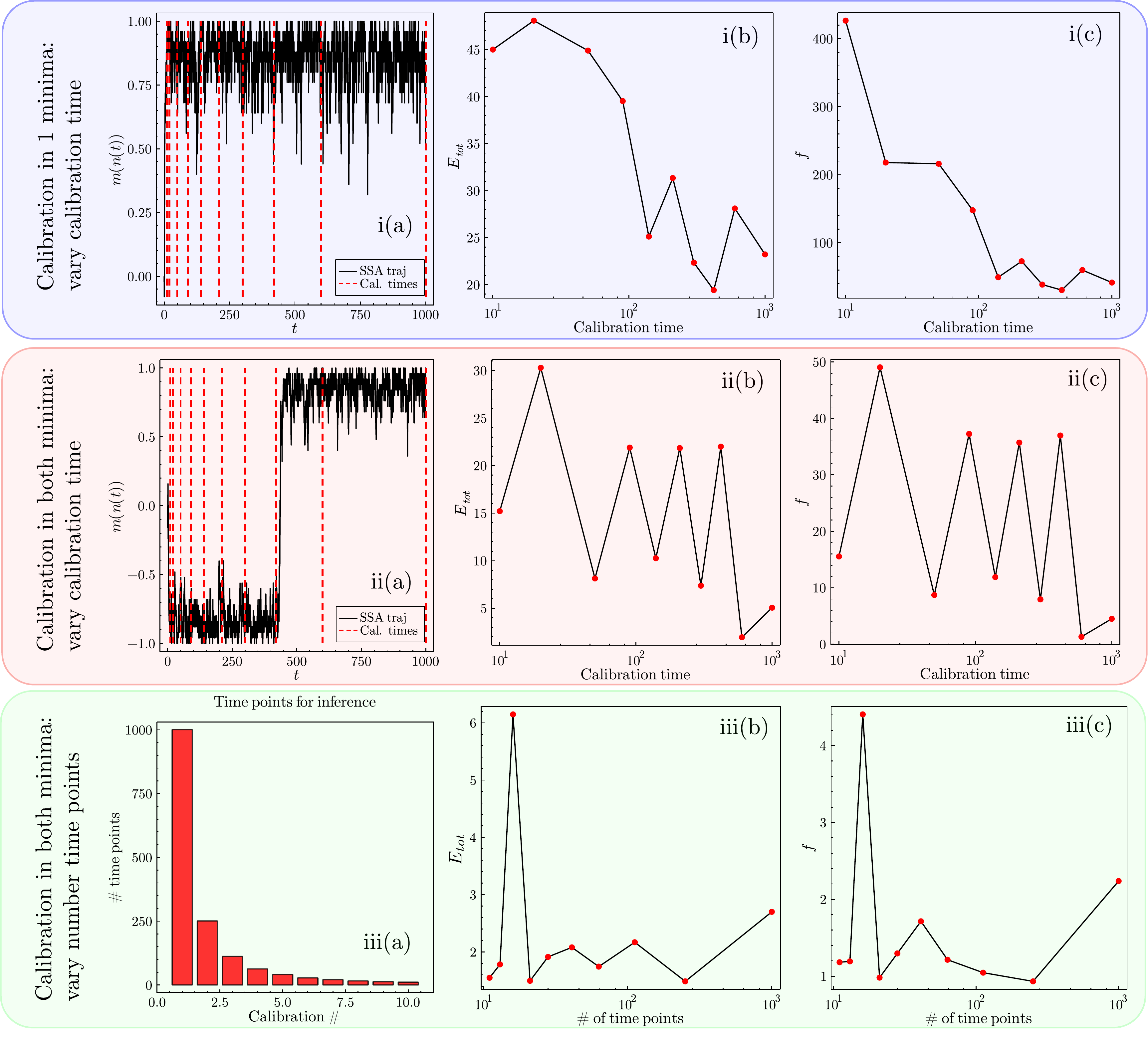}
\caption{Figure showing the calibration procedure on a single trajectory/realisation of data produced by the SSA for the parameter set explored in Fig.~\ref{fig3}. (i) Exploring the calibration when only one equilibrium of the two is realised on the data trajectory. i(a) The data which we perform the calibration on, clearly showing that only the right-hand equilibrium is manifest. Red dashed lines show the various times used for calibration. In i(b) and i(c) we explore the various calibration times and plot the error on the calibrated parameters compared to the true values for $E_{tot}$ and $f$ respectively defined in Eqs.~\eqref{eq:toterr}-\eqref{eq:ferr}. Although larger calibration times result in better parameter inference there are still large errors $\gg 1$ even for large calibration times. (ii) Exploring the calibration procedure for a trajectory that realises both equilibria. ii(a) The trajectory that we use for calibration that explores both equilibria. ii(b) and ii(c) show that increasing the calibration time such that the exploration of both equilibria occurs results in much improved inference of the parameters where for calibration times $\sim 10^3$ the errors are of order $\mathcal{O}(1)$. (iii) Exploring the trajectory in ii(a) for a varied number of time/data points. iii(a) Bar chart showing the different number of time points explored in the calibration. iii(b) and iii(c) show that although having an increased number of time points benefits calibration, it does so with diminishing returns. Note throughout the figure that the optimiser used from \code{BlackBoxOptim} generally identifies good \textit{local minima} of the likelihood function and not the global minimum due to the complexity of the likelihood function \cite{bbobest}.}
\label{fig4}
\end{figure}

First, we look at the case of calibration where only one of the two equilibria is explored, from the trajectory seen in Fig.~\ref{fig4}i(a). The question here is: is there enough information in a single trajectory (from a bimodal system) exploring a single equilibrium mode for the correct parameters to be inferred? The answer is no, which can be seen from a plot of the errors in the inference against the calibration time (for a fixed number of 100 data points) used for the inference in Figs.~\ref{fig4}i(b) and \ref{fig4}i(c). Even though increasing the calibration time improves the calibration, the overall inference of the parameters is still poor for large calibration times with the errors being $\gg 1$. The reason for this is simple: for a system that expresses bimodality the probability is shared across both modes, hence for a bimodal system that only realises one of its two equilibria in a given realisation, the optimiser will instead choose a set of parameters $\theta^\star$ that corresponds to the single mode explored in the data.

Having ruled out the ability to conduct good optimisation when only a single equilibria is explored, we now look at the case where a system realises both of its modes of behaviour on a given trajectory, shown in Fig.~\ref{fig4}ii. The trajectory for this realisation is shown in Fig.~\ref{fig4}ii(a), where the mode of behaviour changes approximately half way through the trajectory. As seen from Fig.~\ref{fig4}ii(b) and \ref{fig4}ii(c), the calibration is significantly improved when both equilibria are explored in the calibration procedure (using a fixed number of 100 data points as before). The reason for this is relatively clear: parameter sets for which the probability distribution exhibits only a single mode near $m=-1$ or $1$ have a lower likelihood, when the system explores the opposite mode from the one they describe. Hence, one can conclude that if the system is bimodal, and the data expresses both of these behaviours, the calibration produced will be relatively good.

Finally, one may want to know the optimal number of data/time points to use, rather than the optimal time of calibration. Fig.~\ref{fig4}iii(a) shows the cases of calibration using differing numbers of data points over the entire trajectory considered in Fig.~\ref{fig4}ii(a), from $1001$ to $10$ data points. Figs.~\ref{fig4}iii(b) and \ref{fig4}ii(c) show the results of inferring the parameters with respect to the errors $E_{tot}$ and $f$: one observes that having more data points is better, but recognises diminishing returns with the additional data points beyond around $10^2$. Therefore, having more data points generally leads to a better parameter estimate but is not as important as having data that represents all behavioural modes of a given parameter set.

\subsection{Calibration from multiple trajectories}\label{sec:multtrajcal}
On some occasions there exist socio-economic binary decisions with more than one realisation. For example, although American presidential elections only happen once every four years, prior to the elections pollsters survey the public to assess what the current opinion is \cite{electionsecon}. \textit{These multiple surveys across the different elections constitute different independent realisations of the same socio-economic phenomena}, if one excludes pollster bias or the possibility that over different elections the underlying model parameters are different. Hence, having multiple realisations of the same phenomena does not mean the existence of multiple realities, but simply temporally or spatially separated independent events which one assumes have similar exogenous and endogenous influences on the agents. Having different realisations can be very beneficial for the calibration procedure, since as we saw for the case of a single trajectory, if only one behavioural mode is explored the calibration is weighted towards parameter sets that favour that reality. We note that the example we explore below is not supposed to represent a very realistic calibration data-set, notably since the initial condition is fixed and the independent trajectories have exactly the same underlying parameter sets (in reality there would be some variations in $F$, $J$ and $\gamma$ between the independent events). However, it shows the considerable effect that multiple realisations have on model calibration. 

\begin{figure}[h!]
\captionsetup{width=1.0\textwidth}
\centering
\includegraphics[width=1.0\textwidth]{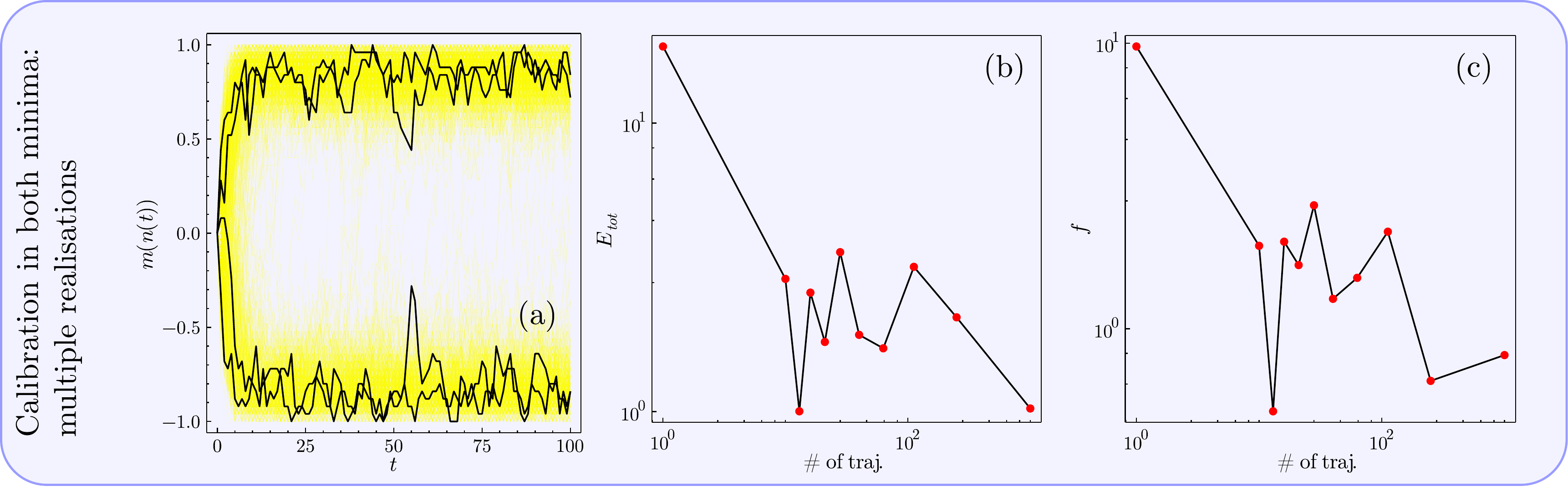}
\caption{Figure showing the performance of the calibration procedure when one has access to multiple realisations of a binary decision process. (a) We simulated $10^3$ trajectories of binary decision data from the SSA for the parameter set used in Fig.~\ref{fig3}. We emphasise five possible trajectories in the foreground, with the yellow background giving an idea of all the trajectories we used. (b) and (c) show that the model calibration can be significantly improved if one has access to multiple realisations of the same binary decision process. For $>10^2$ realisations we find that the errors are typically $\leq1$, which corresponds to a very accurate calibration.}
\label{fig5}
\end{figure}

In Fig.~\ref{fig5}(a) we show 1000 realisations of the process for the parameters in Figs.~\ref{fig3} and \ref{fig4} (in yellow), highlighting five of the trajectories (in black). Figs.~\ref{fig5}(b) and (c) show that for an increased number of realisations the error in the calibration is much reduced, for both $E_{tot}$ and $f$. This reduction in error is even more pronounced than the reduction due to increased calibration times or number of data points seen in Fig.~\ref{fig4}. The reason for this is that if one has multiple trajectories of the same process then it is much more likely that both equilibria will be explored, \textit{and also that the number of trajectories that explore each equilibria is proportional to the probability that a single trajectory will be found in that state}.

In summary, we find that there are several conditions on data that can result in more accurate model calibration, which are:
\begin{enumerate}
    \item[1.] If the system can express bimodal behaviour, and one only has a single data trajectory, then it is imperative that the trajectory used explores both of these behavioural modes. If this is not the case the calibration procedure will weight itself towards parameter sets that only realise the behaviour seen in the data.
    \item[2.] Having a longer calibration time, or an increased number of data points, both improve the accuracy of the calibration, but \textit{with diminishing returns}.
    \item[3.] Having access to more than one realisation of the situation one wishes to model is highly beneficial and results in very accurate calibration. This is seen in Figs.~\ref{fig5}(b) and (c) where the number of trajectories exceed approximately $10^2$. Note also the sizeable drop in error (in log-scale) from a data-set with one realisation to a data-set with 11 realisations in both of these figures. We stress that from a single realisation of real-world data one does not know \textit{a priori} whether the system expresses bimodality, hence having multiple realisations can be very informative, and provide for much better calibration.
\end{enumerate}

\subsection{Other methods for model calibration}
Although it is very convenient for us to utilise the likelihood function based on our time-dependent solution, other likelihood-free calibration methods are available. One of particular mention is \textit{approximate Bayesian computation} (ABC), which generates SSA (Monte Carlo) trajectories for a given parameter set $\theta$ which can then be compared to the original data using a distance function \cite{toni2009approximate,tankhilevich2020gpabc}. The choice of distance function is important, and typical examples include the Euclidean distance between the trajectories, the Hellinger distance, or even the Wasserstein distance \cite{ocal2019parameter}. Often it is best to experiment with each of these, or combinations of them, to see empirically which gives the best result from simulated data, before applying the calibration procedure to a real-world data-set. 

\section{Neoclassical economics versus the binary decision model}\label{sec:compNCE}
In this section we compare the results of the paper to the tenets of neoclassical economics (NCE), and in particular explore: (i) information and expectations, (ii) the dichotomy between the social planner and altruistic agents, (iii) the ideas of equilibrium in economics and complex systems and (iv) the implications of our findings on model calibration to economics. Note that many of the points we make here are not first noted by us, and one may additionally want to read further publications such as \cite{bouchaud2013crises,beinhocker2006origin,kirman1992whom,kirman2010complex,arthur2021foundations,mercure2016modelling,gallegati1999beyond} which have inspired this author.

\subsection{Information and expectations}
NCE assumes that agents have perfect rationality in the sense that all agents can instantaneously perform simultaneous complex calculations and arrive immediately at an equilibrium based on a total knowledge of present and future \cite{beinhocker2006origin}. However, in the binary decision model no two agents make a decision at the same time, and the maximum foresight an agent can have is to choose the decision that maximises the agent's utility \textit{at that point in time}. There is no capacity for the model agents to look into the future, or else to gain an insight into which technology is the optimal one, and hence lock-ins still occur within the binary decision model even where agents are `perfect' (in the limit $\beta\to\infty$). This would not occur in a neoclassical model as the agents would all decide to choose the optimal technology in order to maximise their individual utilities.

\subsection{Social planner versus altruistic agents}
In NCE the ideas of having selfish agents and a social-planner/auctioneer go hand-in-hand. However, in the binary decision model we have explored, the `social planner' situation more closely corresponds to a system of altruistic agents of altruistic strength $\alpha=1$ rather than the system of selfish agents, since the altruistic agents at least consider how each agent can change to best affect the global utility. This dichotomy is curious, since from the perspective of our agent based model the social planner and selfish agents cannot co-exist.

\begin{figure}[h!]
\captionsetup{width=1.0\textwidth}
\centering
\includegraphics[width=0.8\textwidth]{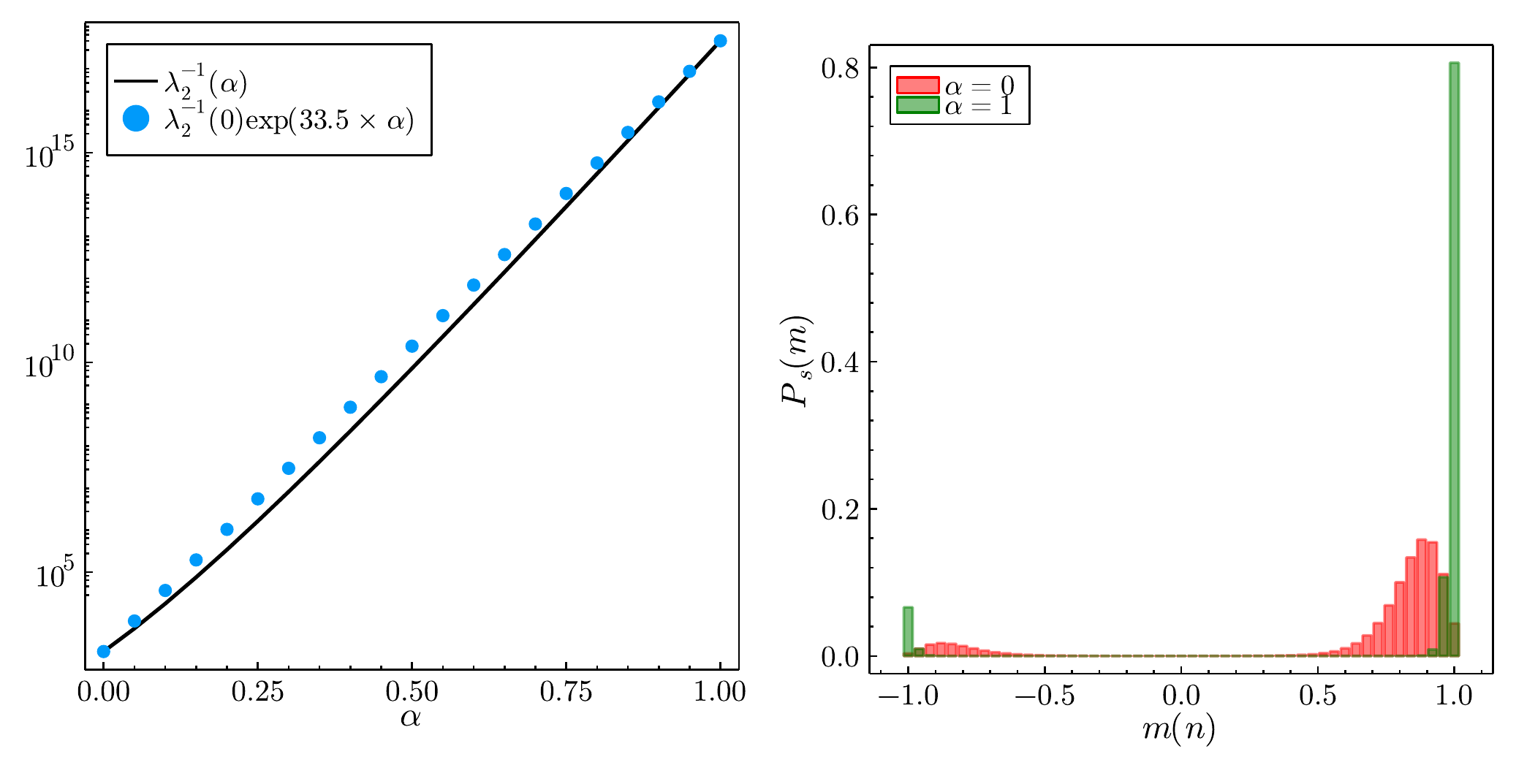}
\caption{For the set of parameters explored in Fig.~\ref{fig3} we explore how the metastable relaxation times (using our very accurate approximation) and the equilibrium distribution behave with respect to changing agent altruism. The left-hand plot shows that as agent altruism increases the metastable waiting times (black line) become exponentially large in $\alpha$. The blue dots show an exponential approximation to $\lambda_2^{-1}$, clearly exhibiting its close-to exponential behaviour. The right-hand plot shows the change in the equilibrium ($t\to\infty$) distributions. For increasing agent altruism populations become more polarised in each behavioural mode (with much reduced fluctuations) even in the metastable regime.}
\label{fig6}
\end{figure}

As Bouchaud noted in \cite{bouchaud2013crises}, the papers of Grauwin et al.~\cite{grauwin2009competition,grauwin2012dynamic} are remarkable in that they exhibit situations in which selfish agents fail spectacularly in a Schelling-like global coordination problem, breaking Adam Smith's `invisible hand', whereas having agents that are even somewhat altruistic remedies the situation for the benefit of all agents. One can further ask a similar question for the mean-field binary decision model: does having altruistic agents increase the ability of the population to increase global utility? The answer to this question is mixed. In one sense the answer is yes---increasing $\alpha$ increases the endogenous influence and hence the time taken to coalesce on a single decision in the population is much reduced. Additionally, in each behavioural mode the fluctuations are reduced if agents are altruistic, shown in the right-hand plot in Fig.~\ref{fig6}, decreasing global utility compared to selfish agents. However, in another sense the answer is no---altruistic agents do not know \textit{a priori} which technology, left or right, is the optimal one and hence can easily coalesce on the wrong technology. In the situation where this occurs the waiting time for them to leave the non-optimal decision is much greater than the selfish agents (shown on the left of Fig.~\ref{fig6}), and in fact the waiting time to equilibrium is roughly exponential in $\alpha$. Hence as Brock and Durlauf \cite{brock2001discrete} state, altruistic agents are more susceptible to conformity effects; \textit{however, their ability to reach the optimal equilibrium mode often does not occur on a relevant time scale.}

\subsection{Equilibrium versus metastability}
The reader may be forgiven if thus far the meaning of equilibrium has become a bit confused, since throughout the paper we have referred to equilibrium, steady-state and metastable states, all of which are distinct from each other. We first note, as pointed out in \cite{beinhocker2006origin}, that both economics and physics share the same definition of equilibrium which is: \textit{a state of balance with no tendency to change} \cite{pass1991harper}. In this paper, the binary choice model that we have looked at is inherently stochastic, and hence we defined the equilibrium with respect to the infinite time limit of the probability distribution in Eq.~\eqref{eq:exactSS}. This means that even at equilibrium there are fluctuations, however small, in the behaviour of the agents. Clearly, this contradicts NCE, in which there is a \textit{single} global optimum.

For this model \textit{equilibrium} and \textit{steady-state} are interchangeable terms, since the model obeys detailed-balance. More generally, \textit{steady-state} even refers to non-detailed balance systems as $t\to\infty$, where there may exist non-equilibrium fluxes in the system \cite{van1992stochastic,tauber2014critical}. \textit{Equilibrium} is a special case of the steady-state, where these fluxes are not present, and the probability flux for microscopic processes in the system is balanced by the reverse process. What we have emphasised in this paper is that where the interaction strengths are strong and agents are sufficiently rational, that metastable states that form have time scales which scale approximately exponentially with the number of agents in the system, and do not relax to the equilibrium in an economically relevant time. \textit{Importantly, the characteristics of these metastable states are highly dependent on the initial condition of the system, hence knowledge of the equilibrium is no longer sufficient.} Such non-equilibrium characteristics cannot be included in the general equilibrium models that economists use so often, presenting a problem for calibration procedures based on the solution of Brock and Durlauf \cite{brock2001discrete}.
\subsection{Implications of calibration procedure}
Previously in Section \ref{sec:multtrajcal}, we defined the conditions for good model calibration, which in essence were: many data points, long calibration times, and multiple trajectories. However, the opinion of some of the literature \cite{sornette2014physics} seems to be that one of the main issues with model calibration lies in not having explicit likelihood functions. In our case, \textit{even with an analytically defined likelihood function, and only three parameters to infer, model calibration is still a difficult task unless one utilises a well-informed data-set.} Additionally, as we have already commented, it is possible to conduct calibration without an explicit likelihood function using likelihood-free methods for parameter inference \cite{tankhilevich2020gpabc}. It is hence vital for model calibration that either calibration methods become much smarter, for example using methods beyond simply using a likelihood function as done in \cite{jia2021frequency} (here in the context of gene expression), or else the data used for calibration becomes more informative. Although it was found that multiple data realisations are the biggest improvement that can be made to the inference procedure, in practise, for many questions of interest, it is not possible to access such data. Hence, the need for quantities of good quality economic data is a pressing one since economists are already using such calibrated models for economic prediction. In lieu of such well informed data-sets with which to conduct econometric analysis, economists are limited to optimising their models for one of many indistinguishable parameter sets from the calibration of their models.

\section{Conclusion}\label{sec:conc}
In this paper we have solved the binary decision model of Brock and Durlauf \cite{brock2001discrete} in time by mapping it to a stochastic birth-death process which can be solved via the method of \cite{smith2015general}. We explored the waiting times to the steady-state in the metastable regime using a very accurate method based on first passage time theory, and used the solution to construct the likelihood function for use in model calibration. The method we employed from \cite{smith2015general} can also be used to solve Kirman's ant recruitment model in time \cite{kirman1993ants,moran2020schrodinger,moran2021ants}. We additionally note that our solution in Eq.~\eqref{eq:exactsol1} is not only a solution to the binary decision model, but is also a time-dependent solution to the mean-field Ising model used as an approximate description of magnetic behaviour in physics. Previous time-dependent solutions to this problem have only come in the form of hydrodynamic approximations \cite{huang1979hydrodynamic}. One of the main utilities of our solution is the ability to infer $\{F,J,\gamma\}$ from real economic binary decision data, and that these three parameters completely specify the dynamics of the agents (up to the choice in decision rule and multiplicative factors $\beta$ and $\alpha$). We constructed a likelihood function and showed that model calibration, even in our simple mean-field model, is a non-trivial task unless one has access to an informative data-set made up of multiple realisations of the same socio-economic phenomena.

Our solution provides a platform for several enhancements and investigations of the model, including:
\begin{enumerate}
    \item[1.] An exploration of non-logit decision rules (such as the Arrhenius form), and additionally decision rules that break detailed balance. This can easily be done using our solution in Eq.~\eqref{eq:exactsol1}, since it is general in the form of the transition rates $W_n(S_i\to-S_i)$. The exploration of non-detailed balance decision rules is of particular importance, as emphasised by Bouchaud \cite{bouchaud2013crises}, since economies are not closed systems but are subject to energy influx/out flow. Recent publications indicate that breaking detailed balance results in system properties that are not consistent with properties of systems pertaining to detailed balance \cite{kumar2020nonequilibrium}.
    \item[2.] In this paper we have not explored our solution with respect to time-dependent $F(t)$ (whose solution is discussed in Appendix \ref{sec:TDZ}), in particular with the rise of metastable states. An understanding of how the time-dependence on $F(t)$ can affect metastability and lock-in effects would have relevance to policy makers, for example in affecting the transition to more renewable energy sources by reducing the time spent fixated in the non-optimal side of a technology lock-in. A question of interest would be: given a system of agents is in the lock-in state in the non-optimal technology, what is the optimal time-dependent function $F(t)$ that releases the agents from this lock-in (in minimal time) whilst keeping $\int_0^{\infty}F(t)dt$ to a minimum (i.e., reducing the effort necessary to break the lock-in)?
    \item[3.] Developing methods to analytically solve multiple choice models, such as those explored in \cite{borghesi2007songs} and \cite{mercure2012ftt}. These models are more generalised than the binary decision model we have discussed in this paper, and in fact contain the binary decision model as a special case, so solutions for multiple choice models would have more potential applications.
    \item[4.] A more detailed investigation of model calibration using more realistic simulated data-sets and real world data-sets. Can one do better than the classic likelihood based inference using more advanced methods? Additionally one could explore the performance of likelihood-free methods of calibration on models for which we do not yet have analytic solutions.
\end{enumerate}
In our opinion, it is the final item in the above list that is of most importance for future investigation, since as we have found even for three parameter inference of $\{F,J,\gamma\}$ in our model, calibration is a difficult task without suitable data-sets. The collection of such data-sets (beyond that of the stock market) that can be used for calibration is therefore of great importance, especially given the need for policy makers to understand immediate challenges relating to the climate crisis and the transition to a more sustainable world \cite{ipcc2021}. As stated by Beinhocker in \textit{The Origin of Wealth}\cite{beinhocker2006origin}, economists will get no sympathy from biologists and physicists who must go to great lengths to collect data to test their theories. Socio-economic modelling, and the models similar to that studied in this paper, can form an essential part of the work necessary to understand what policy makers need to do to change human collective behaviour in uncertain times.

\section*{Acknowledgements}
J.H would like to thank Cambridge Econometrics for hosting him on his NPIF internship, Augustinas Sukys and Pim Vercoulen for thorough proofreading and suggestions, and was supported by a BBSRC EASTBIO PhD studentship.

\section*{Data availability statement}
All code used to make the figures in the paper is available at: \url{https://github.com/jamesholehouse/BinaryDecisionModel}. Similar code instead with application to an Ising magnet is available at: \url{https://github.com/jamesholehouse/Ising}.

\bibliographystyle{naturemag.bst}
\bibliography{main}

\begin{thebibliography}{10}
\expandafter\ifx\csname url\endcsname\relax
  \def\url#1{\texttt{#1}}\fi
\expandafter\ifx\csname urlprefix\endcsname\relax\def\urlprefix{URL }\fi
\providecommand{\bibinfo}[2]{#2}
\providecommand{\eprint}[2][]{\url{#2}}

\bibitem{brock2001discrete}
\bibinfo{author}{Brock, W.~A.} \& \bibinfo{author}{Durlauf, S.~N.}
\newblock \bibinfo{title}{Discrete choice with social interactions}.
\newblock \emph{\bibinfo{journal}{The Review of Economic Studies}}
  \textbf{\bibinfo{volume}{68}}, \bibinfo{pages}{235--260}
  (\bibinfo{year}{2001}).

\bibitem{kirman1993ants}
\bibinfo{author}{Kirman, A.}
\newblock \bibinfo{title}{Ants, rationality, and recruitment}.
\newblock \emph{\bibinfo{journal}{The Quarterly Journal of Economics}}
  \textbf{\bibinfo{volume}{108}}, \bibinfo{pages}{137--156}
  (\bibinfo{year}{1993}).

\bibitem{borghesi2007songs}
\bibinfo{author}{Borghesi, C.} \& \bibinfo{author}{Bouchaud, J.-P.}
\newblock \bibinfo{title}{Of songs and men: a model for multiple choice with
  herding}.
\newblock \emph{\bibinfo{journal}{Quality \& quantity}}
  \textbf{\bibinfo{volume}{41}}, \bibinfo{pages}{557--568}
  (\bibinfo{year}{2007}).

\bibitem{bouchaud2013crises}
\bibinfo{author}{Bouchaud, J.-P.}
\newblock \bibinfo{title}{Crises and collective socio-economic phenomena:
  simple models and challenges}.
\newblock \emph{\bibinfo{journal}{Journal of Statistical Physics}}
  \textbf{\bibinfo{volume}{151}}, \bibinfo{pages}{567--606}
  (\bibinfo{year}{2013}).

\bibitem{hosseiny2019hysteresis}
\bibinfo{author}{Hosseiny, A.}, \bibinfo{author}{Absalan, M.},
  \bibinfo{author}{Sherafati, M.} \& \bibinfo{author}{Gallegati, M.}
\newblock \bibinfo{title}{Hysteresis of economic networks in an xy model}.
\newblock \emph{\bibinfo{journal}{Physica A: Statistical Mechanics and its
  Applications}} \textbf{\bibinfo{volume}{513}}, \bibinfo{pages}{644--652}
  (\bibinfo{year}{2019}).

\bibitem{weisbuch2000market}
\bibinfo{author}{Weisbuch, G.}, \bibinfo{author}{Kirman, A.} \&
  \bibinfo{author}{Herreiner, D.}
\newblock \bibinfo{title}{Market organisation and trading relationships}.
\newblock \emph{\bibinfo{journal}{The economic journal}}
  \textbf{\bibinfo{volume}{110}}, \bibinfo{pages}{411--436}
  (\bibinfo{year}{2000}).

\bibitem{kirman2010complex}
\bibinfo{author}{Kirman, A.}
\newblock \emph{\bibinfo{title}{Complex economics: individual and collective
  rationality}} (\bibinfo{publisher}{Routledge}, \bibinfo{year}{2010}).

\bibitem{moran2021ants}
\bibinfo{author}{Moran, J.}, \bibinfo{author}{Fosset, A.},
  \bibinfo{author}{Kirman, A.} \& \bibinfo{author}{Benzaquen, M.}
\newblock \bibinfo{title}{From ants to fishing vessels: A simple model for
  herding and exploitation of finite resources}.
\newblock \emph{\bibinfo{journal}{Journal of Economic Dynamics and Control}}
  \bibinfo{pages}{104169} (\bibinfo{year}{2021}).

\bibitem{mezard1987spin}
\bibinfo{author}{M{\'e}zard, M.}, \bibinfo{author}{Parisi, G.} \&
  \bibinfo{author}{Virasoro, M.~A.}
\newblock \emph{\bibinfo{title}{Spin glass theory and beyond: An Introduction
  to the Replica Method and Its Applications}}, vol.~\bibinfo{volume}{9}
  (\bibinfo{publisher}{World Scientific Publishing Company},
  \bibinfo{year}{1987}).

\bibitem{moran1958random}
\bibinfo{author}{Moran, P. A.~P.}
\newblock \bibinfo{title}{Random processes in genetics}.
\newblock In \emph{\bibinfo{booktitle}{Mathematical proceedings of the
  cambridge philosophical society}}, vol.~\bibinfo{volume}{54},
  \bibinfo{pages}{60--71} (\bibinfo{organization}{Cambridge University Press},
  \bibinfo{year}{1958}).

\bibitem{moran2020schrodinger}
\bibinfo{author}{Moran, J.}, \bibinfo{author}{Fosset, A.},
  \bibinfo{author}{Benzaquen, M.} \& \bibinfo{author}{Bouchaud, J.-P.}
\newblock \bibinfo{title}{Schr{\"o}dinger’s ants: a continuous description of
  kirman’s recruitment model}.
\newblock \emph{\bibinfo{journal}{Journal of Physics: Complexity}}
  \textbf{\bibinfo{volume}{1}}, \bibinfo{pages}{035002} (\bibinfo{year}{2020}).

\bibitem{farmer2020self}
\bibinfo{author}{Farmer, R.} \& \bibinfo{author}{Bouchaud, J.-P.}
\newblock \bibinfo{title}{Self-fulfilling prophecies, quasi non-ergodicity \&
  wealth inequality}.
\newblock \bibinfo{type}{Tech. Rep.}, \bibinfo{institution}{National Bureau of
  Economic Research} (\bibinfo{year}{2020}).

\bibitem{roberts2002molecular}
\bibinfo{author}{Roberts, K.}, \bibinfo{author}{Alberts, B.},
  \bibinfo{author}{Johnson, A.}, \bibinfo{author}{Walter, P.} \&
  \bibinfo{author}{Hunt, T.}
\newblock \bibinfo{title}{Molecular biology of the cell}.
\newblock \emph{\bibinfo{journal}{New York: Garland Science}}
  (\bibinfo{year}{2002}).

\bibitem{schnoerr2017approximation}
\bibinfo{author}{Schnoerr, D.}, \bibinfo{author}{Sanguinetti, G.} \&
  \bibinfo{author}{Grima, R.}
\newblock \bibinfo{title}{Approximation and inference methods for stochastic
  biochemical kinetics—a tutorial review}.
\newblock \emph{\bibinfo{journal}{Journal of Physics A: Mathematical and
  Theoretical}} \textbf{\bibinfo{volume}{50}}, \bibinfo{pages}{093001}
  (\bibinfo{year}{2017}).

\bibitem{milo2002network}
\bibinfo{author}{Milo, R.} \emph{et~al.}
\newblock \bibinfo{title}{Network motifs: simple building blocks of complex
  networks}.
\newblock \emph{\bibinfo{journal}{Science}} \textbf{\bibinfo{volume}{298}},
  \bibinfo{pages}{824--827} (\bibinfo{year}{2002}).

\bibitem{peccoud1995markovian}
\bibinfo{author}{Peccoud, J.} \& \bibinfo{author}{Ycart, B.}
\newblock \bibinfo{title}{Markovian modeling of gene-product synthesis}.
\newblock \emph{\bibinfo{journal}{Theoretical population biology}}
  \textbf{\bibinfo{volume}{48}}, \bibinfo{pages}{222--234}
  (\bibinfo{year}{1995}).

\bibitem{iyer2009stochasticity}
\bibinfo{author}{Iyer-Biswas, S.}, \bibinfo{author}{Hayot, F.} \&
  \bibinfo{author}{Jayaprakash, C.}
\newblock \bibinfo{title}{Stochasticity of gene products from transcriptional
  pulsing}.
\newblock \emph{\bibinfo{journal}{Physical Review E}}
  \textbf{\bibinfo{volume}{79}}, \bibinfo{pages}{031911}
  (\bibinfo{year}{2009}).

\bibitem{braichenko2021distinguishing}
\bibinfo{author}{Braichenko, S.}, \bibinfo{author}{Holehouse, J.} \&
  \bibinfo{author}{Grima, R.}
\newblock \bibinfo{title}{Distinguishing between models of mammalian gene
  expression: telegraph-like models versus mechanistic models}.
\newblock \emph{\bibinfo{journal}{bioRxiv}}  (\bibinfo{year}{2021}).

\bibitem{schwanhausser2011global}
\bibinfo{author}{Schwanh{\"a}usser, B.} \emph{et~al.}
\newblock \bibinfo{title}{Global quantification of mammalian gene expression
  control}.
\newblock \emph{\bibinfo{journal}{Nature}} \textbf{\bibinfo{volume}{473}},
  \bibinfo{pages}{337--342} (\bibinfo{year}{2011}).

\bibitem{halpern2015bursty}
\bibinfo{author}{Halpern, K.~B.} \emph{et~al.}
\newblock \bibinfo{title}{Bursty gene expression in the intact mammalian
  liver}.
\newblock \emph{\bibinfo{journal}{Molecular cell}}
  \textbf{\bibinfo{volume}{58}}, \bibinfo{pages}{147--156}
  (\bibinfo{year}{2015}).

\bibitem{larsson2019genomic}
\bibinfo{author}{Larsson, A.~J.} \emph{et~al.}
\newblock \bibinfo{title}{Genomic encoding of transcriptional burst kinetics}.
\newblock \emph{\bibinfo{journal}{Nature}} \textbf{\bibinfo{volume}{565}},
  \bibinfo{pages}{251--254} (\bibinfo{year}{2019}).

\bibitem{suter2011mammalian}
\bibinfo{author}{Suter, D.~M.} \emph{et~al.}
\newblock \bibinfo{title}{Mammalian genes are transcribed with widely different
  bursting kinetics}.
\newblock \emph{\bibinfo{journal}{science}} \textbf{\bibinfo{volume}{332}},
  \bibinfo{pages}{472--474} (\bibinfo{year}{2011}).

\bibitem{herbach2017inferring}
\bibinfo{author}{Herbach, U.}, \bibinfo{author}{Bonnaffoux, A.},
  \bibinfo{author}{Espinasse, T.} \& \bibinfo{author}{Gandrillon, O.}
\newblock \bibinfo{title}{Inferring gene regulatory networks from single-cell
  data: a mechanistic approach}.
\newblock \emph{\bibinfo{journal}{BMC systems biology}}
  \textbf{\bibinfo{volume}{11}}, \bibinfo{pages}{1--15} (\bibinfo{year}{2017}).

\bibitem{brock2001interactions}
\bibinfo{author}{Brock, W.~A.} \& \bibinfo{author}{Durlauf, S.~N.}
\newblock \bibinfo{title}{Interactions-based models}.
\newblock In \emph{\bibinfo{booktitle}{Handbook of econometrics}},
  vol.~\bibinfo{volume}{5}, \bibinfo{pages}{3297--3380}
  (\bibinfo{publisher}{Elsevier}, \bibinfo{year}{2001}).

\bibitem{sornette2014physics}
\bibinfo{author}{Sornette, D.}
\newblock \bibinfo{title}{Physics and financial economics (1776--2014):
  puzzles, ising and agent-based models}.
\newblock \emph{\bibinfo{journal}{Reports on progress in physics}}
  \textbf{\bibinfo{volume}{77}}, \bibinfo{pages}{062001}
  (\bibinfo{year}{2014}).

\bibitem{amit1974ginzburg}
\bibinfo{author}{Amit, D.}
\newblock \bibinfo{title}{The ginzburg criterion-rationalized}.
\newblock \emph{\bibinfo{journal}{Journal of Physics C: Solid State Physics}}
  \textbf{\bibinfo{volume}{7}}, \bibinfo{pages}{3369} (\bibinfo{year}{1974}).

\bibitem{bouri2019herding}
\bibinfo{author}{Bouri, E.}, \bibinfo{author}{Gupta, R.} \&
  \bibinfo{author}{Roubaud, D.}
\newblock \bibinfo{title}{Herding behaviour in cryptocurrencies}.
\newblock \emph{\bibinfo{journal}{Finance Research Letters}}
  \textbf{\bibinfo{volume}{29}}, \bibinfo{pages}{216--221}
  (\bibinfo{year}{2019}).

\bibitem{michael2014write}
\bibinfo{author}{Michael, L.} \& \bibinfo{author}{Otterbacher, J.}
\newblock \bibinfo{title}{Write like i write: Herding in the language of online
  reviews}.
\newblock In \emph{\bibinfo{booktitle}{Eighth International AAAI Conference on
  Weblogs and Social Media}} (\bibinfo{year}{2014}).

\bibitem{cornwall1989determinants}
\bibinfo{author}{Cornwall, M.}
\newblock \bibinfo{title}{The determinants of religious behavior: A theoretical
  model and empirical test}.
\newblock \emph{\bibinfo{journal}{Social forces}}
  \textbf{\bibinfo{volume}{68}}, \bibinfo{pages}{572--592}
  (\bibinfo{year}{1989}).

\bibitem{tauber2014critical}
\bibinfo{author}{T{\"a}uber, U.~C.}
\newblock \emph{\bibinfo{title}{Critical dynamics: a field theory approach to
  equilibrium and non-equilibrium scaling behavior}}
  (\bibinfo{publisher}{Cambridge University Press}, \bibinfo{year}{2014}).

\bibitem{van1992stochastic}
\bibinfo{author}{Van~Kampen, N.~G.}
\newblock \emph{\bibinfo{title}{Stochastic processes in physics and
  chemistry}}, vol.~\bibinfo{volume}{1} (\bibinfo{publisher}{Elsevier},
  \bibinfo{year}{1992}).

\bibitem{glauber1963time}
\bibinfo{author}{Glauber, R.~J.}
\newblock \bibinfo{title}{Time-dependent statistics of the ising model}.
\newblock \emph{\bibinfo{journal}{Journal of mathematical physics}}
  \textbf{\bibinfo{volume}{4}}, \bibinfo{pages}{294--307}
  (\bibinfo{year}{1963}).

\bibitem{nadal1998formal}
\bibinfo{author}{Nadal, J.-P.}, \bibinfo{author}{Weisbuch, G.},
  \bibinfo{author}{Chenevez, O.} \& \bibinfo{author}{Kirman, A.}
\newblock \bibinfo{title}{A formal approach to market organization: choice
  functions, mean field approximation and maximum entropy principle}.
\newblock \emph{\bibinfo{journal}{Advances in Self-Organization and
  Evolutionary Economics}} \bibinfo{pages}{149--159} (\bibinfo{year}{1998}).

\bibitem{mercure2012ftt}
\bibinfo{author}{Mercure, J.-F.}
\newblock \bibinfo{title}{{FTT}: Power: A global model of the power sector with
  induced technological change and natural resource depletion}.
\newblock \emph{\bibinfo{journal}{Energy Policy}}
  \textbf{\bibinfo{volume}{48}}, \bibinfo{pages}{799--811}
  (\bibinfo{year}{2012}).

\bibitem{mercure2011global}
\bibinfo{author}{Mercure, J.-F.}
\newblock \bibinfo{title}{Global electricity technology substitution model with
  induced technological change}.
\newblock \emph{\bibinfo{journal}{Tyndall Centre on Global Climate Change
  Working Paper}}  (\bibinfo{year}{2011}).

\bibitem{anderson1992discrete}
\bibinfo{author}{Anderson, S.~P.}, \bibinfo{author}{De~Palma, A.} \&
  \bibinfo{author}{Thisse, J.-F.}
\newblock \emph{\bibinfo{title}{Discrete choice theory of product
  differentiation}} (\bibinfo{publisher}{MIT press}, \bibinfo{year}{1992}).

\bibitem{mercure2016modelling}
\bibinfo{author}{Mercure, J.-F.}, \bibinfo{author}{Pollitt, H.},
  \bibinfo{author}{Bassi, A.~M.}, \bibinfo{author}{Vi{\~n}uales, J.~E.} \&
  \bibinfo{author}{Edwards, N.~R.}
\newblock \bibinfo{title}{Modelling complex systems of heterogeneous agents to
  better design sustainability transitions policy}.
\newblock \emph{\bibinfo{journal}{Global environmental change}}
  \textbf{\bibinfo{volume}{37}}, \bibinfo{pages}{102--115}
  (\bibinfo{year}{2016}).

\bibitem{mccullen2013multiparameter}
\bibinfo{author}{McCullen, N.~J.}, \bibinfo{author}{Rucklidge, A.~M.},
  \bibinfo{author}{Bale, C.~S.}, \bibinfo{author}{Foxon, T.~J.} \&
  \bibinfo{author}{Gale, W.~F.}
\newblock \bibinfo{title}{Multiparameter models of innovation diffusion on
  complex networks}.
\newblock \emph{\bibinfo{journal}{SIAM Journal on Applied Dynamical Systems}}
  \textbf{\bibinfo{volume}{12}}, \bibinfo{pages}{515--532}
  (\bibinfo{year}{2013}).

\bibitem{beinhocker2006origin}
\bibinfo{author}{Beinhocker, E.~D.}
\newblock \emph{\bibinfo{title}{The origin of wealth: Evolution, complexity,
  and the radical remaking of economics}} (\bibinfo{publisher}{Harvard Business
  Press}, \bibinfo{year}{2006}).

\bibitem{king2015advanced}
\bibinfo{author}{King, J.~E.}
\newblock \emph{\bibinfo{title}{Advanced introduction to post Keynesian
  economics}} (\bibinfo{publisher}{Edward Elgar Publishing},
  \bibinfo{year}{2015}).

\bibitem{gillespie2007stochastic}
\bibinfo{author}{Gillespie, D.~T.}
\newblock \bibinfo{title}{Stochastic simulation of chemical kinetics}.
\newblock \emph{\bibinfo{journal}{Annu. Rev. Phys. Chem.}}
  \textbf{\bibinfo{volume}{58}}, \bibinfo{pages}{35--55}
  (\bibinfo{year}{2007}).

\bibitem{grauwin2009competition}
\bibinfo{author}{Grauwin, S.}, \bibinfo{author}{Bertin, E.},
  \bibinfo{author}{Lemoy, R.} \& \bibinfo{author}{Jensen, P.}
\newblock \bibinfo{title}{Competition between collective and individual
  dynamics}.
\newblock \emph{\bibinfo{journal}{Proceedings of the National Academy of
  Sciences}} \textbf{\bibinfo{volume}{106}}, \bibinfo{pages}{20622--20626}
  (\bibinfo{year}{2009}).

\bibitem{gardiner2009stochastic}
\bibinfo{author}{Gardiner, C.}
\newblock \emph{\bibinfo{title}{Stochastic methods}}, vol.~\bibinfo{volume}{4}
  (\bibinfo{publisher}{Springer Berlin}, \bibinfo{year}{2009}).

\bibitem{smith2015general}
\bibinfo{author}{Smith, S.} \& \bibinfo{author}{Shahrezaei, V.}
\newblock \bibinfo{title}{General transient solution of the one-step master
  equation in one dimension}.
\newblock \emph{\bibinfo{journal}{Physical Review E}}
  \textbf{\bibinfo{volume}{91}}, \bibinfo{pages}{062119}
  (\bibinfo{year}{2015}).

\bibitem{pillai2005perron}
\bibinfo{author}{Pillai, S.~U.}, \bibinfo{author}{Suel, T.} \&
  \bibinfo{author}{Cha, S.}
\newblock \bibinfo{title}{The perron-frobenius theorem: some of its
  applications}.
\newblock \emph{\bibinfo{journal}{IEEE Signal Processing Magazine}}
  \textbf{\bibinfo{volume}{22}}, \bibinfo{pages}{62--75}
  (\bibinfo{year}{2005}).

\bibitem{weidlich1971statistical}
\bibinfo{author}{Weidlich, W.}
\newblock \bibinfo{title}{The statistical description of polarization phenomena
  in society}.
\newblock \emph{\bibinfo{journal}{British Journal of Mathematical and
  Statistical Psychology}} \textbf{\bibinfo{volume}{24}},
  \bibinfo{pages}{251--266} (\bibinfo{year}{1971}).

\bibitem{bezanson2017julia}
\bibinfo{author}{Bezanson, J.}, \bibinfo{author}{Edelman, A.},
  \bibinfo{author}{Karpinski, S.} \& \bibinfo{author}{Shah, V.~B.}
\newblock \bibinfo{title}{Julia: A fresh approach to numerical computing}.
\newblock \emph{\bibinfo{journal}{SIAM review}} \textbf{\bibinfo{volume}{59}},
  \bibinfo{pages}{65--98} (\bibinfo{year}{2017}).
\newblock \urlprefix\url{https://doi.org/10.1137/141000671}.

\bibitem{munsky2006finite}
\bibinfo{author}{Munsky, B.} \& \bibinfo{author}{Khammash, M.}
\newblock \bibinfo{title}{The finite state projection algorithm for the
  solution of the chemical master equation}.
\newblock \emph{\bibinfo{journal}{The Journal of chemical physics}}
  \textbf{\bibinfo{volume}{124}}, \bibinfo{pages}{044104}
  (\bibinfo{year}{2006}).

\bibitem{iserles2019applications}
\bibinfo{author}{Iserles, A.} \& \bibinfo{author}{MacNamara, S.}
\newblock \bibinfo{title}{Applications of magnus expansions and pseudospectra
  to markov processes}.
\newblock \emph{\bibinfo{journal}{European Journal of Applied Mathematics}}
  \textbf{\bibinfo{volume}{30}}, \bibinfo{pages}{400--425}
  (\bibinfo{year}{2019}).

\bibitem{tapias2020entropic}
\bibinfo{author}{Tapias, D.}, \bibinfo{author}{Paprotzki, E.} \&
  \bibinfo{author}{Sollich, P.}
\newblock \bibinfo{title}{From entropic to energetic barriers in glassy
  dynamics: The barrat--m{\'e}zard trap model on sparse networks}.
\newblock \emph{\bibinfo{journal}{Journal of Statistical Mechanics: Theory and
  Experiment}} \textbf{\bibinfo{volume}{2020}}, \bibinfo{pages}{093302}
  (\bibinfo{year}{2020}).

\bibitem{schnakenberg1976network}
\bibinfo{author}{Schnakenberg, J.}
\newblock \bibinfo{title}{Network theory of microscopic and macroscopic
  behavior of master equation systems}.
\newblock \emph{\bibinfo{journal}{Reviews of Modern physics}}
  \textbf{\bibinfo{volume}{48}}, \bibinfo{pages}{571} (\bibinfo{year}{1976}).

\bibitem{ferrenberg2018pushing}
\bibinfo{author}{Ferrenberg, A.~M.}, \bibinfo{author}{Xu, J.} \&
  \bibinfo{author}{Landau, D.~P.}
\newblock \bibinfo{title}{Pushing the limits of monte carlo simulations for the
  three-dimensional ising model}.
\newblock \emph{\bibinfo{journal}{Physical Review E}}
  \textbf{\bibinfo{volume}{97}}, \bibinfo{pages}{043301}
  (\bibinfo{year}{2018}).

\bibitem{morningstar2017deep}
\bibinfo{author}{Morningstar, A.} \& \bibinfo{author}{Melko, R.~G.}
\newblock \bibinfo{title}{Deep learning the ising model near criticality}.
\newblock \emph{\bibinfo{journal}{arXiv preprint arXiv:1708.04622}}
  (\bibinfo{year}{2017}).

\bibitem{cervera2018exact}
\bibinfo{author}{Cervera-Lierta, A.}
\newblock \bibinfo{title}{Exact ising model simulation on a quantum computer}.
\newblock \emph{\bibinfo{journal}{Quantum}} \textbf{\bibinfo{volume}{2}},
  \bibinfo{pages}{114} (\bibinfo{year}{2018}).

\bibitem{horst2008ergodicity}
\bibinfo{author}{Horst, U.}
\newblock \bibinfo{title}{Ergodicity and non-ergodicity in economics}.
\newblock \emph{\bibinfo{journal}{New Palgrave Dictionary of Economics, revised
  edition, L. Blume and S. Durlauf, eds}}  (\bibinfo{year}{2008}).

\bibitem{nakicenovic1996freeing}
\bibinfo{author}{Naki{\'c}enovi{\'c}, N.}
\newblock \bibinfo{title}{Freeing energy from carbon}.
\newblock \emph{\bibinfo{journal}{Daedalus}} \textbf{\bibinfo{volume}{125}},
  \bibinfo{pages}{95--112} (\bibinfo{year}{1996}).

\bibitem{farmer2016predictable}
\bibinfo{author}{Farmer, J.~D.} \& \bibinfo{author}{Lafond, F.}
\newblock \bibinfo{title}{How predictable is technological progress?}
\newblock \emph{\bibinfo{journal}{Research Policy}}
  \textbf{\bibinfo{volume}{45}}, \bibinfo{pages}{647--665}
  (\bibinfo{year}{2016}).

\bibitem{foxon2002technological}
\bibinfo{author}{Foxon, T.~J.}
\newblock \bibinfo{title}{Technological and institutional ‘lock-in’as a
  barrier to sustainable innovation}.
\newblock \emph{\bibinfo{journal}{Imperial College Centre for Policy and
  Technology Working Paper}}  (\bibinfo{year}{2002}).

\bibitem{ashcroft2016metastable}
\bibinfo{author}{Ashcroft, P.}
\newblock \bibinfo{title}{Metastable states in a model of cancer initiation}.
\newblock In \emph{\bibinfo{booktitle}{The Statistical Physics of Fixation and
  Equilibration in Individual-Based Models}}, \bibinfo{pages}{91--126}
  (\bibinfo{publisher}{Springer}, \bibinfo{year}{2016}).

\bibitem{antal2006fixation}
\bibinfo{author}{Antal, T.} \& \bibinfo{author}{Scheuring, I.}
\newblock \bibinfo{title}{Fixation of strategies for an evolutionary game in
  finite populations}.
\newblock \emph{\bibinfo{journal}{Bulletin of mathematical biology}}
  \textbf{\bibinfo{volume}{68}}, \bibinfo{pages}{1923--1944}
  (\bibinfo{year}{2006}).

\bibitem{qin2005self}
\bibinfo{author}{Qin, A.~K.} \& \bibinfo{author}{Suganthan, P.~N.}
\newblock \bibinfo{title}{Self-adaptive differential evolution algorithm for
  numerical optimization}.
\newblock In \emph{\bibinfo{booktitle}{2005 IEEE congress on evolutionary
  computation}}, vol.~\bibinfo{volume}{2}, \bibinfo{pages}{1785--1791}
  (\bibinfo{organization}{IEEE}, \bibinfo{year}{2005}).

\bibitem{bbobest}
\bibinfo{author}{Feldt, R.}
\newblock \bibinfo{title}{\code{BlackBoxOptim}}.
\newblock
  \bibinfo{howpublished}{\url{https://github.com/robertfeldt/BlackBoxOptim.jl/blob/master/examples/benchmarking/latest_toplist.csv}}
  (\bibinfo{year}{2021}).
\newblock \bibinfo{note}{[Online; accessed 12-August-2021]}.

\bibitem{horst2000introduction}
\bibinfo{author}{Horst, R.}, \bibinfo{author}{Pardalos, P.~M.} \&
  \bibinfo{author}{Van~Thoai, N.}
\newblock \emph{\bibinfo{title}{Introduction to global optimization}}
  (\bibinfo{publisher}{Springer Science \& Business Media},
  \bibinfo{year}{2000}).

\bibitem{electionsecon}
\bibinfo{author}{{The Economist}}.
\newblock \bibinfo{title}{{Forecasting the US elections}}.
\newblock
  \bibinfo{howpublished}{\url{https://projects.economist.com/us-2020-forecast/president}}
  (\bibinfo{year}{2020}).
\newblock \bibinfo{note}{[Online; accessed 14-August-2021]}.

\bibitem{toni2009approximate}
\bibinfo{author}{Toni, T.}, \bibinfo{author}{Welch, D.},
  \bibinfo{author}{Strelkowa, N.}, \bibinfo{author}{Ipsen, A.} \&
  \bibinfo{author}{Stumpf, M.~P.}
\newblock \bibinfo{title}{Approximate bayesian computation scheme for parameter
  inference and model selection in dynamical systems}.
\newblock \emph{\bibinfo{journal}{Journal of the Royal Society Interface}}
  \textbf{\bibinfo{volume}{6}}, \bibinfo{pages}{187--202}
  (\bibinfo{year}{2009}).

\bibitem{tankhilevich2020gpabc}
\bibinfo{author}{Tankhilevich, E.} \emph{et~al.}
\newblock \bibinfo{title}{Gpabc: a julia package for approximate bayesian
  computation with gaussian process emulation}.
\newblock \emph{\bibinfo{journal}{Bioinformatics}}
  \textbf{\bibinfo{volume}{36}}, \bibinfo{pages}{3286--3287}
  (\bibinfo{year}{2020}).

\bibitem{ocal2019parameter}
\bibinfo{author}{{\"O}cal, K.}, \bibinfo{author}{Grima, R.} \&
  \bibinfo{author}{Sanguinetti, G.}
\newblock \bibinfo{title}{Parameter estimation for biochemical reaction
  networks using wasserstein distances}.
\newblock \emph{\bibinfo{journal}{Journal of Physics A: Mathematical and
  Theoretical}} \textbf{\bibinfo{volume}{53}}, \bibinfo{pages}{034002}
  (\bibinfo{year}{2019}).

\bibitem{kirman1992whom}
\bibinfo{author}{Kirman, A.~P.}
\newblock \bibinfo{title}{Whom or what does the representative individual
  represent?}
\newblock \emph{\bibinfo{journal}{Journal of economic perspectives}}
  \textbf{\bibinfo{volume}{6}}, \bibinfo{pages}{117--136}
  (\bibinfo{year}{1992}).

\bibitem{arthur2021foundations}
\bibinfo{author}{Arthur, W.~B.}
\newblock \bibinfo{title}{Foundations of complexity economics}.
\newblock \emph{\bibinfo{journal}{Nature Reviews Physics}}
  \textbf{\bibinfo{volume}{3}}, \bibinfo{pages}{136--145}
  (\bibinfo{year}{2021}).

\bibitem{gallegati1999beyond}
\bibinfo{author}{Gallegati, M.} \& \bibinfo{author}{Kirman, A.}
\newblock \emph{\bibinfo{title}{Beyond the representative agent}}
  (\bibinfo{publisher}{Edward Elgar Publishing}, \bibinfo{year}{1999}).

\bibitem{grauwin2012dynamic}
\bibinfo{author}{Grauwin, S.}, \bibinfo{author}{Goffette-Nagot, F.} \&
  \bibinfo{author}{Jensen, P.}
\newblock \bibinfo{title}{Dynamic models of residential segregation: An
  analytical solution}.
\newblock \emph{\bibinfo{journal}{Journal of Public Economics}}
  \textbf{\bibinfo{volume}{96}}, \bibinfo{pages}{124--141}
  (\bibinfo{year}{2012}).

\bibitem{pass1991harper}
\bibinfo{author}{Pass, C.}, \bibinfo{author}{Lowes, B.},
  \bibinfo{author}{Davis, L.} \& \bibinfo{author}{Kronish, S.~J.}
\newblock \bibinfo{title}{The harper collins dictionary of economics}
  (\bibinfo{year}{1991}).

\bibitem{jia2021frequency}
\bibinfo{author}{Jia, C.} \& \bibinfo{author}{Grima, R.}
\newblock \bibinfo{title}{Frequency domain analysis of fluctuations of mrna and
  protein copy numbers within a cell lineage: theory and experimental
  validation}.
\newblock \emph{\bibinfo{journal}{Physical Review X}}
  \textbf{\bibinfo{volume}{11}}, \bibinfo{pages}{021032}
  (\bibinfo{year}{2021}).

\bibitem{huang1979hydrodynamic}
\bibinfo{author}{Huang, H.}
\newblock \bibinfo{title}{Hydrodynamic solution of the time-dependent ising
  model}.
\newblock \emph{\bibinfo{journal}{The Journal of Chemical Physics}}
  \textbf{\bibinfo{volume}{70}}, \bibinfo{pages}{2390--2392}
  (\bibinfo{year}{1979}).

\bibitem{kumar2020nonequilibrium}
\bibinfo{author}{Kumar, M.} \& \bibinfo{author}{Dasgupta, C.}
\newblock \bibinfo{title}{Nonequilibrium phase transition in an ising model
  without detailed balance}.
\newblock \emph{\bibinfo{journal}{Physical Review E}}
  \textbf{\bibinfo{volume}{102}}, \bibinfo{pages}{052111}
  (\bibinfo{year}{2020}).

\bibitem{ipcc2021}
\bibinfo{author}{{Masson-Delmotte, V., P. Zhai, A. Pirani, S. L. Connors, C.
  Péan, S. Berger, N. Caud, Y. Chen, L. Goldfarb, M. I. Gomis, M. Huang, K.
  Leitzell, E. Lonnoy, J.B.R. Matthews, T. K. Maycock, T. Waterfield, O.
  Yelekçi, R. Yu and B. Zhou (eds.)}}.
\newblock \emph{\bibinfo{title}{{IPCC, 2021: Summary for Policymakers. In:
  \textit{Climate Change 2021: The Physical Science Basis. Contribution of
  Working Group I to the Sixth Assessment Report of the Intergovernmental Panel
  on Climate Change}}}} (\bibinfo{publisher}{Cambridge University Press. In
  Press}).

\bibitem{cao2020analytical}
\bibinfo{author}{Cao, Z.} \& \bibinfo{author}{Grima, R.}
\newblock \bibinfo{title}{Analytical distributions for detailed models of
  stochastic gene expression in eukaryotic cells}.
\newblock \emph{\bibinfo{journal}{Proceedings of the National Academy of
  Sciences}} \textbf{\bibinfo{volume}{117}}, \bibinfo{pages}{4682--4692}
  (\bibinfo{year}{2020}).

\bibitem{gillespie1977exact}
\bibinfo{author}{Gillespie, D.~T.}
\newblock \bibinfo{title}{Exact stochastic simulation of coupled chemical
  reactions}.
\newblock \emph{\bibinfo{journal}{The journal of physical chemistry}}
  \textbf{\bibinfo{volume}{81}}, \bibinfo{pages}{2340--2361}
  (\bibinfo{year}{1977}).

\bibitem{anderson2007modified}
\bibinfo{author}{Anderson, D.~F.}
\newblock \bibinfo{title}{A modified next reaction method for simulating
  chemical systems with time dependent propensities and delays}.
\newblock \emph{\bibinfo{journal}{The Journal of chemical physics}}
  \textbf{\bibinfo{volume}{127}}, \bibinfo{pages}{214107}
  (\bibinfo{year}{2007}).

\bibitem{thanh2015simulation}
\bibinfo{author}{Thanh, V.~H.} \& \bibinfo{author}{Priami, C.}
\newblock \bibinfo{title}{Simulation of biochemical reactions with
  time-dependent rates by the rejection-based algorithm}.
\newblock \emph{\bibinfo{journal}{The Journal of chemical physics}}
  \textbf{\bibinfo{volume}{143}}, \bibinfo{pages}{08B601\_1}
  (\bibinfo{year}{2015}).

\bibitem{arfken1999mathematical}
\bibinfo{author}{Arfken, G.~B.} \& \bibinfo{author}{Weber, H.~J.}
\newblock \bibinfo{title}{Mathematical methods for physicists}
  (\bibinfo{year}{1999}).

\end{thebibliography}

\appendix

\section{1D master equation}\label{sec:mastereqn}
In this section we derive the master equation for a single discrete stochastic variable $n$. There are many sources for this, but the primary sources we use for problems of this type are \cite{van1992stochastic,gardiner2009stochastic}. First we introduce $W_{nn'}$ as the transition probability per unit time to transition from state $n'$ to $n$, and hence the probability to transition from $n'$ to $n$ in a time interval $\Delta t$ is $W_{nn'}\Delta t$. One can then write an equation for the conservation of probability at $t+\Delta t$,
\begin{align}
    \mathcal{P}(n,t+\Delta t) = \sum_{n'\neq n}\mathcal{P}(n',t)W_{nn'}\Delta t + \left(1-\sum_{n'\neq n}W_{n'n}\Delta t\right)\mathcal{P}(n,t).
\end{align}
The first term on the right-hand side of this equation describes the flow of probability into the state $n$ from all other states $n'$ in $[t,t+\Delta t)$, whereas the second term describes the probability of remaining in state $n$ in $[t,t+\Delta t)$. One can rearrange this, taking $\mathcal{P}(n,t)$ on the right-hand side to the left, and dividing everything by $\Delta t$. In the limit $\Delta t\to 0$ we then arrive at the 1D master equation,
\begin{align}\label{eq:genME}
    \partial_t \mathcal{P}(n,t) = \sum_{n'\neq n}W_{nn'}\mathcal{P}(n',t)-\mathcal{P}(n,t)\sum_{n'\neq n}W_{n'n},
\end{align}
subject to the initial condition $\mathcal{P}(n,0) = \mathcal{Q}(n)$ for some general initial distribution $\mathcal{Q}(n)$ and the normalisation condition $\sum_n \mathcal{P}(n,t) = 1$. From the form of Eq.~\eqref{eq:genME} one can see the master equation as a `gain-loss' equation in the probability $\mathcal{P}(n,t)$, the first right-hand side term describing the gain and the second describing the loss \cite{van1992stochastic}. For the derivation of Eq.~\eqref{eq:me} in the main text we identify,
\begin{align}
    W_{n n'} = \begin{cases}
      b_{n}, & n'=n+1, \\
      a_{n}, & n'=n-1, \\
      0, & \text{otherwise}.
   \end{cases}
\end{align}

\section{Time-dependent zeitgeist solution}\label{sec:TDZ}
In this section we derive the analytic solution to the mean-field binary decision model under time-dependent $F(t)$. Let $F(t)$ be a piecewise defined function up to the time of interest $t_T$, divided over $T$ intervals of equal length, where each interval is constant, i.e., $F(t_{j-1}<t<t_j) = F_j$ for $j\in\{1,2,...,T\}$ and $t_0 = 0$. In the case where $F(t)$ is not piecewise defined it can be approximated to any degree of accuracy as a piecewise function where $F_j\approx F(t_{j-1})$ and the accuracy of the approximation is improved for increasing $T$. Then over each time interval the propensities $a^j_i(t)$ and $b^j_i(t)$ for interval $j$ take constant values and hence the master operator in Eq.~\eqref{eq:TRM} is a constant matrix within each interval. The solution requires that in each of these intervals that the eigenvalues $\lambda_i^{j}$ for interval $j$ and $i\in\{1,2,...,N+1\}$ are calculated computationally. Note that the more piecewise elements of $F(t)$ there are the longer this will take computationally. For the first interval, given some initial condition $\mathbb{Q}_1(m(n))$ at $t = 0$, the solution in $0<t<t_1$ is,
\begin{align}
    P_1(m(n),0\leq t<t_1|\mathbb{Q}_1(m(n)),0) = \sum_{n_0 = 0}^N \mathbb{Q}_1(m(n_0))P_1(m(n),t|m(n_0),0)
\end{align}
where $P_1(m(n),t|m(n_0),0)$ is defined as in Eqs.~(\ref{eq:exactsol1}-\ref{eq:exactsolend}) but with $\lambda_i\to \lambda_i^1$. For the next interval $t_1<t<t_2$ the solution is very similar but we now use the initial condition $\mathbb{Q}_2(m(n)) = P_1(m(n),t_1|\mathbb{Q}_1(m(n)),0)$, giving us,
\begin{align}
    P_2(m(n),t_1\leq t<t_2|\mathbb{Q}_2(m(n)),t_1) = \sum_{n_0 = 0}^N \mathbb{Q}_2(m(n_0))P_2(m(n),t-t_1|m(n_0),0),
\end{align}
where one now uses $\lambda_i\to \lambda_i^2$ in Eqs.~(\ref{eq:exactsol1}-\ref{eq:exactsolend}). One then repeats this process for all intervals $j\in\{3,4,...,T\}$ up to $P_T(m(n),t_{T-1}<t<t_T|\mathbb{Q}_{T-1}(m(n)),t_{T-1})$. This completes the solution. We note that this is a similar method to how time-dependent transcription rates are dealt with in Section 10 of the supplementary information of \cite{cao2020analytical}.

\section{Stochastic simulation algorithm}\label{sec:SSA}
The SSA, also known as the \textit{Gillespie algorithm} named after the scientist who popularised its use, provides a popular Monte Carlo method to simulate the economic model considered in this paper \cite{gillespie1977exact,gillespie2007stochastic}. The major benefit of the SSA is that, unlike the Metropolis-Hastings algorithm \cite{tauber2014critical} or the Glauber update algorithm \cite{glauber1963time}, the SSA provides a continuous time description of stochastic processes. In this paper we use the SSA to simulate stochastic trajectories for systems of economic agents based on the transition probabilities for each agent to change their decision in Eq.~\eqref{eq:genrule}. Below we will re-introduce some of the formulae previously presented in the paper for the reader's benefit. Note that although we use the SSA in this context to simulate trajectories from the mean-field binary decision model, one can use the same method for the more generalised system of non mean-field agents, agents with personal preferences described at the start of Section \ref{sec:econ_agents}.

Consider a system of mean-field economic agents described in Section \ref{sec:econ_agents}. A given agent $i$ at a time $t$ will have made a decision $S_i \in \{-1,1\}$, and at some future point in the future the agent can change their decision to $-S_i$. The rate at which any agent $i$ will change their decision is given by,
\begin{align}
    W_n(S_i\to -S_i) = \frac{\gamma}{1+\exp(-\beta \mathcal{G}_i)},
\end{align}
which is dependent on the number of agents $n$ already deciding for $S_i = 1$ and the gain function defined in Eq.~\eqref{eq:gain}. Now, since all the agents have the same influence $I(n,t)$ upon them, we can further define the total propensities with which any agent can change their decision from $1\to-1$ or $-1\to 1$, respectively,
\begin{align}\nonumber
    f^+_n &= (N-n)W_n(-1\to 1),\\\nonumber
    f^-_n &= n W_n(1\to -1).
\end{align}
These propensities follow intuitively from the law of mass-action since the rate at which agents change their decision from $1\to-1$ is proportional to the number of right deciding agents $n$ \cite{gillespie2007stochastic,schnoerr2017approximation}. We denote the total propensity at which any decision changes are made as $f_n = f^+_n+f^-_n$, and state the \textit{fundamental premise} that in time interval $[t,t+\Delta t)$ the probability that \textit{any} agent will change their decision is $f_n \Delta t$ \cite{gillespie2007stochastic}. It follows that agent decision changes are exponentially distributed, and the waiting time $u$ for the next change of decision (of any type) given the current state of the system $n$ is drawn from,
\begin{align}
    u \sim \frac{1}{f_n}\exp(-f_n u).
\end{align}
The probability that this decision change at $t+u$ is of type is $1\to-1$ is $f_n^-/f_n$, and the probability that this decision change is $-1\to1$ is $f_n^+/f_n$. One then samples which decision will change next from these probabilities and updates the number of left and right voting agents accordingly. \textit{Note that this algorithm can only be used in the current form (and via the direct method below) where $F(t)$ does not have time dependence, in which case, modifications to the algorithm must be considered} \cite{anderson2007modified,thanh2015simulation}. We now detail the pseudo-code for the SSA algorithm via the \textit{direct method} (a computationally faster but identical approach to the more intuitive algorithm above) \cite{gillespie2007stochastic}:
\begin{enumerate}
    \item[1.] For $t=0$, initialise the state variable of the system defined by the number of right voting agents $n$ in a population of $N$ agents. As each agent has a randomised initial choice with probability $p$ of deciding right, $n$ is drawn from a binomial distribution, i.e., $n\sim \text{Bin}(N,p)$ (other initial conditions can be used). Define a $M+1$ element $\vec{V}$ vector that stores the state of the system at times $\mathcal{T}=\{0, \delta t, 2 \delta t,... ,M \delta t\}$, for some time step $\delta t$. Set $V_0 = n$.
    \item[2.] Calculate $f_n$ for the current state of the system $n$.
    \item[3.] Draw 2 random numbers, $r_1$ and $r_2$, from the uniform distribution over the unit interval $[0,1]$. The waiting time, $u$, for the next decision change is then found to be,
    $$ u = \frac{1}{f_n}\ln\left(\frac{1}{r_1}\right).$$
    \item[4.] Update the storage of the state of the system between $[t,t+u)$. For every element of index $i$ in $\mathcal{T}$ such that $t\leq \mathcal{T}_i \leq t+u$ assign $V_i = n$. Then, update the system time as $t \to t+u$. If $f_n^+>r_2 f_n$ then flip the decision of left voting agent to right, i.e., update $n \to n+1$, otherwise flip the decision of right voting left to right, i.e., update $n \to n-1$.
    \item[5.] If $t > M \delta t$ stop the simulation and return $\vec{V}$, otherwise go back to step $2$.
\end{enumerate}
Steps 1-5 detail the SSA for a single trajectory. In order to calculate the probability state vector, mean and variance at the times $\mathcal{T} = \{0, \delta t, 2 \delta t,... ,M \delta t\}$ an ensemble of simulations would need to be produced. Say one produces an ensemble of $E$ simulations each of which outputs a state vector $\vec{n}_i,\; i\in [1,2,...,E]$ of length $M+1$, where $\vec{n}_i$ is the state vector over all times in $\mathcal{T}$ for ensemble simulation $i$. Note that due to the stochastic nature of the SSA, each trajectory in the ensemble will be different from each other \textit{even when the initial conditions of each simulation are the same.} The mean and variance at each time point, over the ensemble of simulations is given by,
\begin{align}
    \langle n((m-1)\delta t)\rangle &= \frac{1}{E}\sum_{i=1}^E [n_i]_m,\\
    \text{Var}(n((m-1)\delta t)) &= \frac{1}{E}\sum_{i=1}^E [n_i]_m^2 - \langle n((m-1)\delta t)\rangle^2,
\end{align}
for $m\in[1,2,...,M+1]$, where $[n_i]_m$ is the $m^{\text{th}}$ measurement of the $i^{\text{th}}$ trajectory. The probability distribution $\mathcal{P}(n,t)$ is instead given by the normalised histogram over each time slice in the ensemble. Explicitly this is,
\begin{align}
    \mathcal{P}(n,(m-1)\delta t) = \frac{\text{\# of times $n$ appears in $[n_i]_m,\; \forall \;i\in[1,2,...,E]$}}{E}.
\end{align}
One can verify that for each slice of time the probability distribution is indeed normalised, i.e., $\sum_n \mathcal{P}(n,t) = 1$.

\begin{figure}[h!]
\captionsetup{width=1.0\textwidth}
\centering
\includegraphics[width=0.6\textwidth]{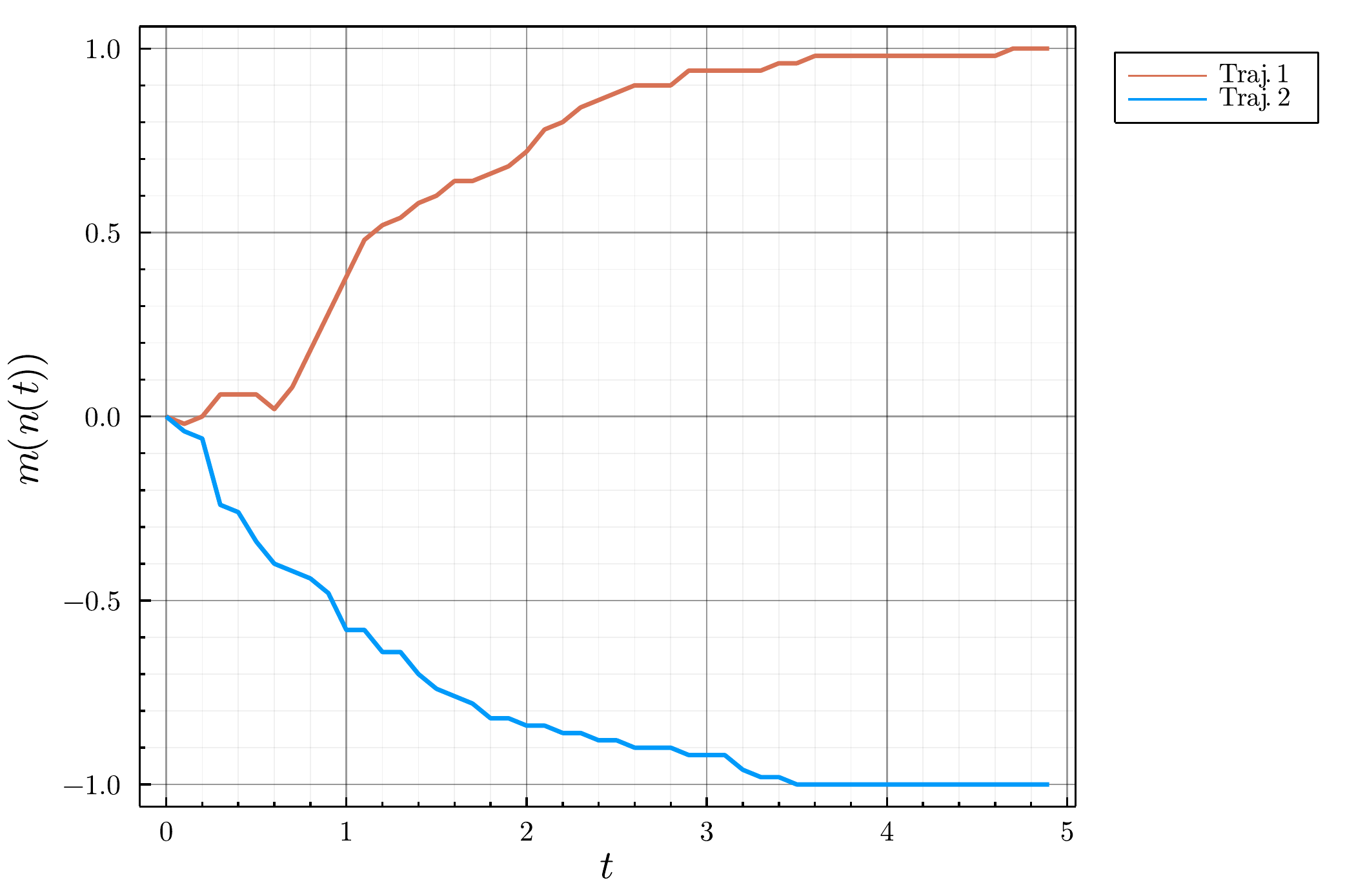}
\caption{For the parameters shown in Fig.~ \ref{fig1}(c) in the main text there are two associated trajectories that the system can undertake. Either the SSA trajectory will show a steady evolution to $m(N)=1$ or else it will evolve to $m(0)=-1$. These two realisations of the process occur with the same probability for $F=0$ and $J>1/\beta(\alpha+1)$.}
\label{supp1}
\end{figure}

\section{Criticality in the mean-field model}\label{sec:criticality}
\begin{figure}[h!]
\captionsetup{width=1.0\textwidth}
\centering
\includegraphics[width=1.0\textwidth]{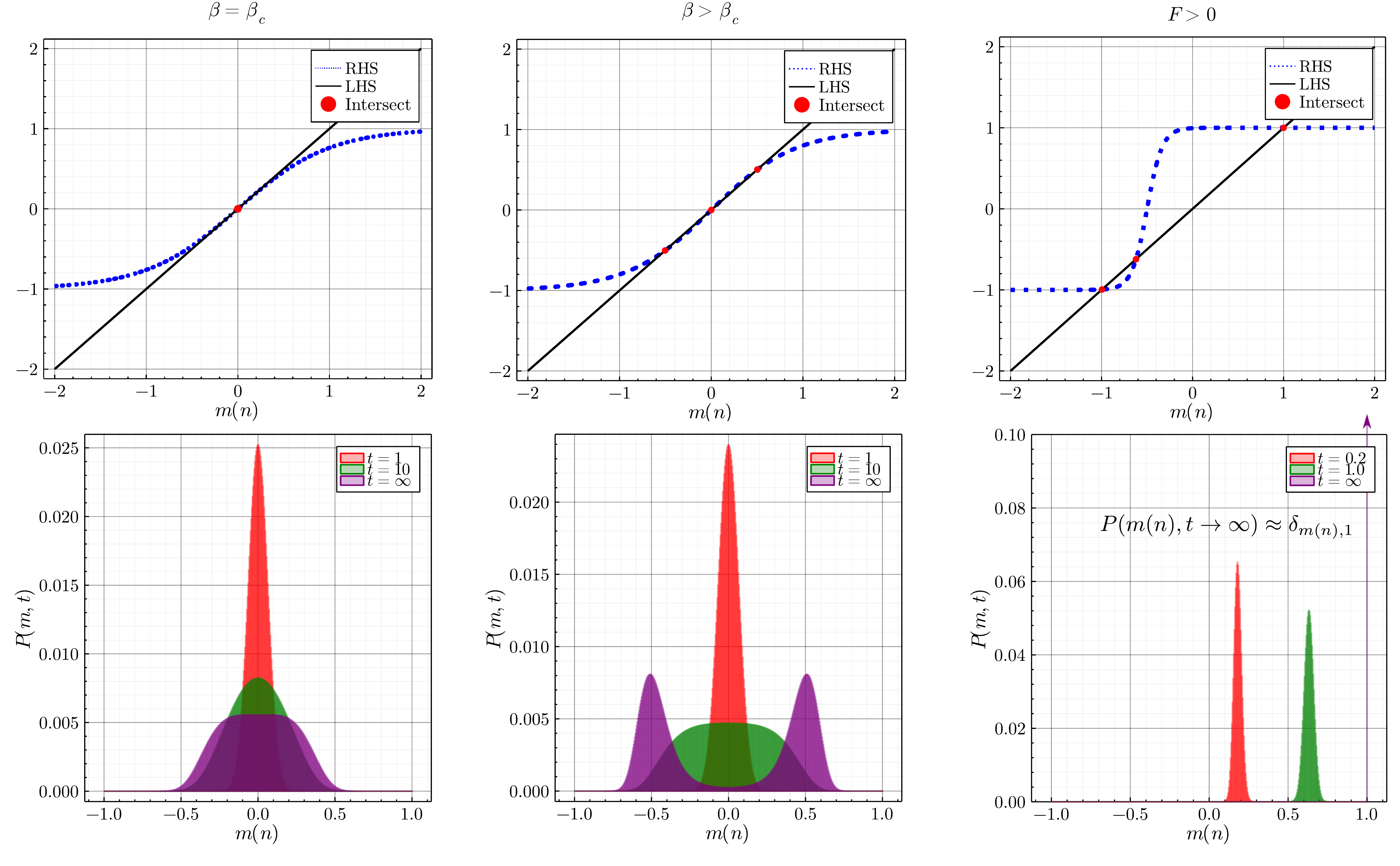}
\caption{Plots show the critical behaviour of the mean-field model for $N = 500$. The top row of plots shows the solutions of the equation $m = \tanh(\beta(F+J(\alpha+1)m))$ as red points for different values of $\beta$. LHS denotes the left-hand side of this equation and RHS denotes the right-hand side. The bottom row of plots shows the evolution of the probability distributions over $m(n)$ from near the initial condition to the steady-state, obtained from Eqs.~(\ref{eq:exactsol1}) and (\ref{eq:exactSS}), and correspond to plots in the row above them. Parameters for column 1 are: $J=1,\alpha=0,F=0$ and $\beta=1$. Parameters for column 2 are: $J=1,\alpha=0,F=0$ and $\beta=1.1$. Parameters for column 3 are: $J=1,\alpha=0,F=0.5$ and $\beta=6$.}
\label{supp2}
\end{figure}

In this section we explore some interesting features of the mean-field model relating to critical behaviour. For $F=0$ this allows us to identify a critical rationality of the agents $\beta_c$ where for any $\beta>\beta_c$ we observe the occurrence of a bimodal steady-state distribution for $m(n)$. First though, it is useful to have a deterministic description of the dynamics such that one can identify the possible equilibrium values of $m$ for a given parameter set. From the dynamics of \eqref{eq:rs} we write the deterministic rate equation as follows,
\begin{align}\label{eq:detRE}
    \partial_t \langle n\rangle = (N-\langle n\rangle) r(\langle n\rangle )-\langle n\rangle l(\langle n\rangle ),
\end{align}
where $r(n)$ and $l(n)$ are rates defined in the main text. Solving this at steady-state in the limit $N\to\infty$ one can show,
\begin{align}\label{eq:transcend}
    m = \tanh\left(\beta(F+J(1+\alpha)m)\right)
\end{align}
This equation is transcendental and one must identify the solutions to it computationally. It is interesting to ask how many solutions of $m$ one expects for different values of the parameters. Consider setting the zeitgeist to zero, $F=0$, where the only influence on the agents is now from the mean-field interactions. For there to be 3 solutions (2 stable, 1 unstable determined by the Jacobian of Eq.~\eqref{eq:detRE}) of this equation one sees that the gradient of the right-hand side of Eq.~\eqref{eq:transcend} with respect to $m$ must be greater than 1. To leading order in an expansion in $m$, one can identify a critical value of agent rationality, $\beta_c = 1/J(\alpha+1)$, for which any small increase above $\beta_c$ leads to 3 solutions of Eq.~\eqref{eq:transcend} and two stable equilibrium values of $m$. One can see this from the top left and central plots in Fig.~\ref{supp2}, for $\beta=\beta_c$ there is 1 intersection point between the two sides of the equation, whereas for $\beta\gtrsim\beta_c$ there are 3 intersection points, with the middle intersection point being unstable. This is exhibited in the analytical solution from Eq.~\eqref{eq:exactsol1} in the bottom left and central plots: for $\beta=\beta_c$ the steady-state distribution has a very flat top and any small increase in $\beta$ leads to a bimodal steady-state distribution. This bimodal behaviour is seen in individual realisations of the SSA in Fig.~\ref{supp1}: for $\beta>\beta_c$ agent behaviour takes one of two directions, they either drift to mostly deciding right \textit{or} left.

Finally, one can consider what happens for $F\neq 0$, which is explored in the final column of Fig.~\ref{supp2}. In some situations we find three solutions to Eq.~\eqref{eq:transcend}, in other cases only 1. However, even where 3 solutions are found (as in Fig.~\ref{supp2}) the agents are exponentially more likely to favour being in the configuration where their decision has the same sign as the sign of $F$. This is seen in the bottom right plot in Fig.~\ref{supp2} for $F>0$, even though there exists a second stable solution at $m\approx -1$, as $t\to\infty$ the steady-state is $P(m,t\to\infty)\approx \delta_{m,1}.$ This exponential behaviour is further supported by Eqs.~(\ref{eq:approx_tau1}-\ref{eq:approx_tau2}) in the main text.

\section{Derivation of $\phi_R$}\label{sec:FP}
In this section we derive $\phi_R$, the probability to end up with the agents coalescing on the right-hand technology given they start at $n_u$ (or equivalently $m_u$ with respect to the order parameter). Note that, for the sake of repetition the reader may already want to have read Section \ref{sec:exactdet} in the main text, since we reference necessary equations from that section in the derivation below. We calculate $\phi_R$ in the same way that fixation probabilities are calculated for birth-death processes with two absorbing boundaries (see \cite{van1992stochastic,ashcroft2016metastable}). As stated in the main text, we calculate the `equilibrium values' of $n$ based on the extrema of the steady-state probability distribution. In the case where we satisfy the conditions set out in Section \ref{sec:metastab} and Appendix \ref{sec:bouchaud_calc} ($\beta>\beta_c$ and $|F|<J(1+\alpha)$) there will be three extremal values, the middle one corresponding to the unstable equilibrium $n_u$ and the other two, $n_-< n_u$ and $n_+>n_u$, being the stable modes of the bimodal agent behaviour (i.e., the stable equilibrium points).

\begin{figure}[h!]
\captionsetup{width=1.0\textwidth}
\centering
\includegraphics[width=0.6\textwidth]{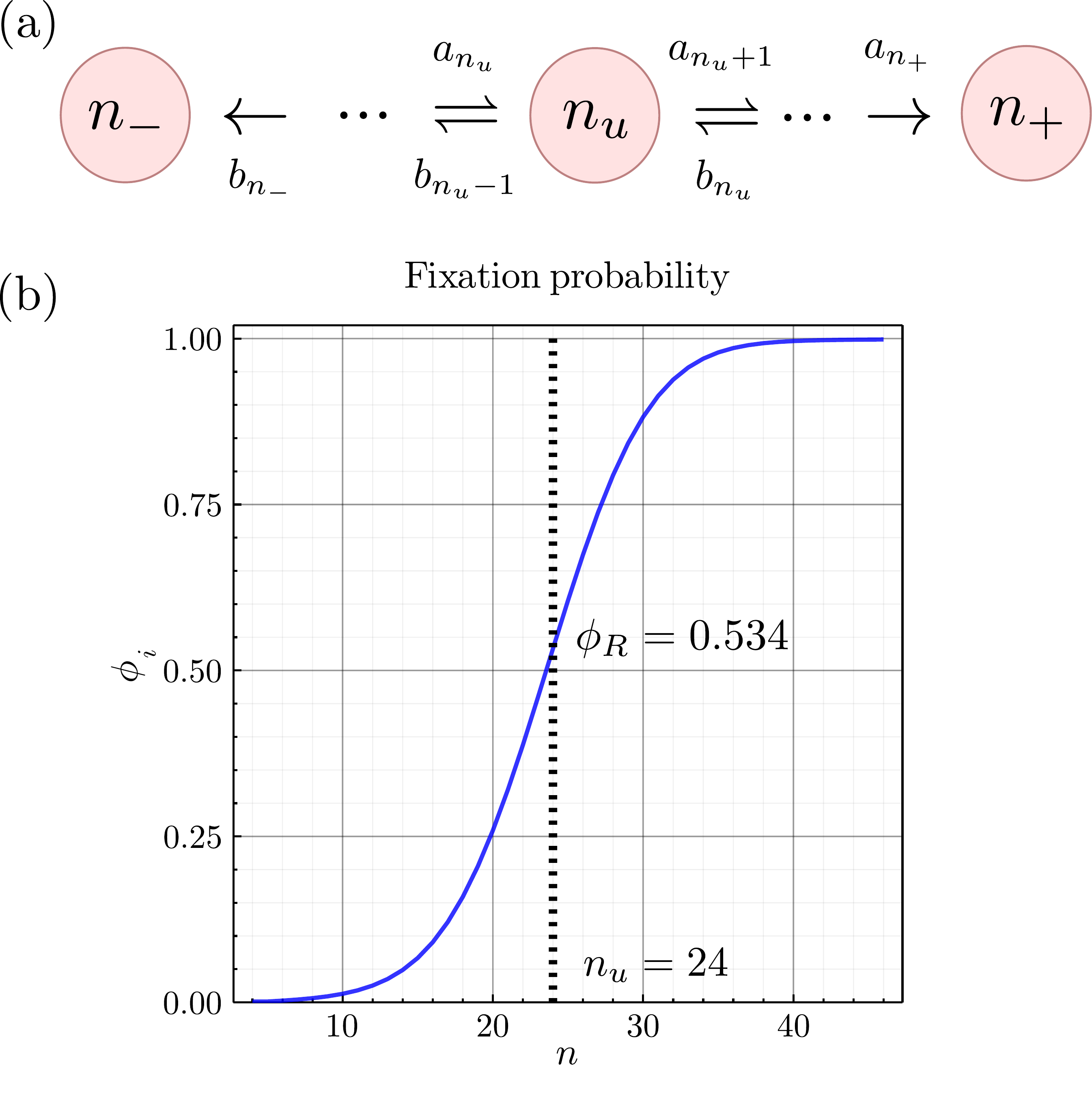}
\caption{Plots for the calculation of $\phi_R$. (a) We calculate $\phi_R$ as a fixation probability to end up with a system of agent coalesced on the right technology by artificially imposing absorbing boundaries at the modes of the behaviour of the agents. From this artificial system $\phi_R$ is well defined. (b) Plot of the fixation probability for the same parameters as Fig.~\ref{fig3}. For this parameter set we find the equilibria values from the steady-state distribution are $n_- = 3$, $n_+ = 47$ and $n_u = 24$ (noting $N=50$ and $F>0$). Clearly since $F>0$ it makes sense that at the unstable equilibrium there is a slightly greater chance of end up at the right technology which is optimal for the agents.}
\label{supp3}
\end{figure}

Now for the calculation of $\phi_R$ we must consider a separate microstate diagram to the one explored in Fig.~\ref{fig1}(b), in particular $n_-$ and $n_+$ become absorbing states, as shown in Fig.~\ref{supp3}(a). The question we now ask is given that one starts at some $n$ in $n_-<n<n_+$, what is the probability $\phi_i$ of getting fixated at the right-hand mode? Note that in our notation $\phi_R \equiv \phi_{n_u}$. The value of $1-\phi_R$ then gives the probability of getting fixated at the left-hand mode. To proceed we see that $\phi_i = \lim_{t\to\infty}Q_{n_+,i}(t)$ from Eq.~\eqref{eq:bme}, i.e., the probability of being found at $n=n_+$ as $t\to\infty$, given one start at $t=0$ at $n=i$, for the microscopic transitions in Fig.~\ref{supp3}(a). Hence, we get the following recursive equation for $\phi_i$,
\begin{align}
    \phi_i = a_{i+1} \Delta t \phi_i + b_{i-1} \Delta t \phi_{i-1} + (1-(a_{i+1}+b_{i-1})\Delta t)\phi_i,
\end{align}
where we note that by definition $\phi_{n_-} = 0$ and $\phi_{n_+} = 1$. To solve this equation we introduce the difference variable $\nu_i = \phi_i-\phi_{i-1}$, which converts this equation into,
\begin{align}
    \nu_i = \frac{b_{i-2}}{a_{i}}\nu_{i-1}.
\end{align}
Solving this equation recursively then gives,
\begin{align}
    \nu_i = \left(\prod_{j=n_-+1}^{i-1}\frac{b_{i-1}}{a_{i+1}}\right)\phi_{n_-+1},
\end{align}
since $\nu_{n_-+1} = \phi_{n_-+1}$. In order to find $\phi_{n_-+1}$ one can then show that $\sum_{k=n_-+1}^{n_+} \nu_k = \phi_{n_+} = 1$ and hence we find,
\begin{align}
    \phi_{n_-+1} = \left(1 + \sum_{k=n_-+1}^{n_+-1} \prod_{j=n_-+1}^{k}\frac{b_{i-1}}{a_{i+1}} \right)^{-1}.
\end{align}
Finally, we can calculate all the $\phi_i$ through,
\begin{align}
    \phi_i = \sum_{k=n_-+1}^i \nu_k = \frac{1 + \sum_{k=n_-+1}^{i-1} \prod_{j=n_-+1}^{k}\frac{b_{i-1}}{a_{i+1}}}{1 + \sum_{k=n_-+1}^{n_+-1} \prod_{j=n_-+1}^{k}\frac{b_{i-1}}{a_{i+1}}}.
\end{align}
This completes our derivation of $\phi_R = \phi_{n_u}$. In Fig.~\ref{supp3}(b) we plot the fixation probability $\phi_i$ for the parameters of Fig.~\ref{fig3} in the main text, and find $\phi_u \sim 0.534$. This is expected, since $F>0$ there a greater chance of first hitting the right boundary than the left one.

\section{Derivation of approximate relaxation time scales}\label{sec:bouchaud_calc}
In this section we derive Eqs.~\eqref{eq:approx_tau1}-\eqref{eq:approx_tau2} in the main text. Following the notation of \cite{bouchaud2013crises} we define $\phi = n/N = (m+1)/2$, which is the fraction of agents choosing the right technology. Using the results from the standard textbooks \cite{van1992stochastic,gardiner2009stochastic}, one can approximate the master equation in Eq.~\eqref{eq:me} by a Fokker-Planck equation (FPE),
\begin{align}
    \partial_t \mathcal{P}(n,t) = -\partial_n[((N-n)r(n)-n l(n))\mathcal{P}(n,t)]+\frac{1}{2}\partial_n^2[((N-n)r(n)+n l(n))\mathcal{P}(n,t)],
\end{align}
where $n$ is now a continuous variable $\in(-\infty,\infty)$. The FPE is a good approximation of the original master equation if $N\gg 1$ and $r(n)$ and $l(n)$ are smooth, slowly varying functions with respect to continuous $n$. One can then change variables from $n\to \phi$ to arrive at the following FPE now in $\phi$,
\begin{align}\label{eq:FPEphi}
    \partial_t \mathcal{P}(\phi,t) = -\partial_\phi (a_1(\phi)\mathcal{P}(\phi,t))+\frac{1}{2N}\partial_\phi^2(a_2(\phi)\mathcal{P}(\phi,t)),
\end{align}
where $\partial_n = N^{-1}\partial_\phi$ and we have defined,
\begin{align}
    a_1(\phi) &= \gamma\left((1-\phi)-\left(1+e^{\beta f(\phi)}\right)^{-1}\right),\\
    a_2(\phi) &= \gamma\left((1-\phi)+(2\phi-1)\left(1+e^{\beta f(\phi)}\right)^{-1}\right),
\end{align}
with,
\begin{align}
    f(\phi) = 2(F+J(\alpha+1)(2\phi-1)).
\end{align}
Note that since $N\gg1$ we have ignored the negligible self interaction term. In the limit $N\to \infty$ the noise term disappears entirely and the FPE equation reduces to a deterministic description via a rate equation in the mean value $\langle\phi\rangle$. Multiplying Eq.~\eqref{eq:FPEphi} by $\phi$, setting the noise term to zero and integrating over all $\phi$ we arrive at the rate equation,
\begin{align}
    \partial_t \langle\phi\rangle = \langle a_1(\phi)\rangle \approx a_1(\langle\phi\rangle),
\end{align}
with the steady-state value(s) of $\phi$ determined by $a_1(\langle\phi\rangle_s) = 0$. Note that the approximation $\langle a_1(\phi)\rangle \approx a_1(\langle\phi\rangle)$ is valid so long as the fluctuations of $\phi$ about $\langle \phi\rangle$ are small compared to $\langle \phi\rangle$. For $\beta>\beta_c$ and $|F|<J(1+\alpha)$ there are three equilibrium solutions of $\langle\phi\rangle_s$: two of which are stable $\langle\phi\rangle_s = 0,1$ and one of which is unstable $\langle\phi\rangle_u=\frac{1}{2}(1-\frac{F}{J(\alpha+1)})$. If either of these conditions are broken then there is only one equilibrium solution. Our interest now is to find mean first passage times of getting to the unstable equilibrium point, starting from each of the stable equilibria. 

Again, following standard methods for the FPE \cite{van1992stochastic}, one can find the mean first passage time to reach $\phi_u$ given initially at $\phi$ as,
\begin{align}\label{eq:approxtau1}
    \tau_\phi &= 2 N\begin{cases} 
      \int_\phi^{\phi_u}e^{\Phi(y')}dy'\int_0^{y'}e^{-\Phi(y'')}\frac{ dy''}{a_2(y'')}, \; & \phi<\phi_u,\\
      \int_{\phi_u}^{\phi}e^{\Phi(y')}dy'\int_{y'}^{1}e^{-\Phi(y'')}\frac{ dy''}{a_2(y'')}, \; & \phi>\phi_u.
   \end{cases}
\end{align}
where we have further defined,
\begin{align}\label{eq:Phidef}
    \Phi(y) = -2N\int_0^{y}\frac{a_1(y)}{a_2(y)}dy.
\end{align}
In order to proceed we investigate the limit $\beta \to \infty$, i.e., where agents are highly rational. In this limit one can show that,
\begin{align}
    \frac{a_1(y)}{a_2(y)} \sim
    \begin{cases} 
      -1, & y>\phi_u, \\
      0, & y=\phi_u, \\
      1, & y < \phi_u.
   \end{cases}
\end{align}
This then gives us a simplified form of $\Phi(y)$ in the highly rational limit,
\begin{align}
    \Phi(y) \sim
    \begin{cases} 
      2Ny, & y\leq\phi_u, \\
      2N(2\phi_u-y), & y\geq\phi_u.
   \end{cases}
\end{align}
$\Phi(y)$ is now approximately a triangular function with its maximum at $\phi_u$ and minima at 0 and 1. Since the function is peaked around $\phi_u$ one can then successively use two saddle point approximations on Eqs.~\eqref{eq:approxtau1} \cite{arfken1999mathematical}. Doing so gives our required mean first passage times in the limit $\beta\to \infty$,
\begin{align}
    \tau_{lr} &= \tau_0 \sim \frac{2\pi}{\sqrt{\Phi''(0)|\Phi''(\phi_u)|}}\exp\left( N\left(1-\frac{F}{J\left(\alpha+1\right)}\right) \right),\\
    \tau_{rl} &= \tau_1 \sim \frac{2\pi}{\sqrt{\Phi''(1)|\Phi''(\phi_u)|}}\exp\left( N\left(1+\frac{F}{J\left(\alpha+1\right)}\right) \right),
\end{align}
where the double prime denotes the second derivative of $\Phi(y)$ from Eq.~\eqref{eq:Phidef}. Note that shown here is the full result of the calculation, whereas in the main text we show only the proportionality to the exponential function. As one would expect from symmetry, if $F=0$ then $\tau_{lr} = \tau_{rl}$  since $\Phi(y)$ becomes a symmetric function about $\phi_u$.

\end{document}